\documentclass[journal]{IEEEtran}
\usepackage{xcolor,soul,framed}
\usepackage{ifpdf}
\usepackage{float}
\usepackage{cite}
\ifCLASSINFOpdf
\usepackage[pdftex]{graphicx}
\DeclareGraphicsExtensions{.pdf,.jpeg,.png}
\else
\usepackage[dvips]{graphicx}
\DeclareGraphicsExtensions{.pdf}
\fi
\usepackage[caption=false, font=normalsize]{subfig}

\usepackage{amsthm} 
\usepackage{balance}
\usepackage{textcomp}
\usepackage{epstopdf}
\usepackage{amsmath}\DeclareMathOperator*{\argmax}{argmax}
\usepackage{caption}
\usepackage{amssymb}
\usepackage{bbm}
\usepackage{color}
\usepackage{balance}
\usepackage{array}
\newcolumntype{P}[1]{>{\centering\arraybackslash}p{#1}}
\usepackage{mathtools} 

\usepackage[ruled,vlined,linesnumbered]{algorithm2e}
\usepackage{enumitem} 
\usepackage{esvect} 
\usepackage{tabularx} 
\usepackage{multirow} 
\usepackage{booktabs} 
\usepackage{makecell} 
\usepackage{aligned-overset} 
\usepackage{url}
\usepackage{multicol} 
\usepackage{xcolor}
\usepackage{mathtools} 

\graphicspath{{./Figures/}}
\newtheorem{Remark}{Remark}

\makeatletter
\renewcommand*\env@matrix[1][*\l@MaxMatrixCols l]{%
 \hskip -\arraycolsep
 \let\@ifnextchar\new@ifnextchar
 \array{#1}}
\makeatother

\renewcommand{\vec}[1]{\boldsymbol{\mathrm{#1}}}

\newcommand{\removelatexerror}{\let\@latex@error\@gobble}

\begin{document}

\title{DRL-Based Maximization of the Sum Cross-Layer Achievable Rate for Networks Under Jamming}
   \author{
  {Abdul Basit\IEEEauthorrefmark{1}, Muddasir Rahim\IEEEauthorrefmark{1}, Tri~Nhu~Do\IEEEauthorrefmark{2}, Nadir~Adam\IEEEauthorrefmark{1}, and Georges~Kaddoum\IEEEauthorrefmark{1}~\IEEEmembership{Senior~Member,~IEEE}}
  
\thanks{\IEEEauthorrefmark{1}A.~Basit,~M.~Rahim,~N.~Adam,~and~G.~Kaddoum are with the Department of Electrial Engineering, \'{E}cole de Technologie Sup\'{e}rieure (\'{E}TS), Universit\'{e} du Qu\'{e}bec, Montr\'{e}al, QC, Canada (emails:~abdul.basit.1@ens.etsmtl.ca, muddasir.rahim.1@ens.etsmtl.ca, nadir.adam@etsmtl.ca, georges.kaddoum@etsmtl.ca).
 
  \IEEEauthorrefmark{2}T.~N.~Do is with the Department of Electrical Engineering, Polytechnique Montreal, QC, Canada (email: tri-nhu.do@polymtl.ca).
  }}
\maketitle
\begin{abstract}
In quasi-static wireless networks characterized by infrequent changes in the transmission schedules of user equipment (UE), malicious jammers can easily deteriorate network performance. Accordingly, a key challenge in these networks is managing channel access amidst jammers and under dynamic channel conditions. In this context, we propose a robust learning-based mechanism for channel access in multi-cell quasi-static networks under jamming. 
The network comprises multiple legitimate UEs, including predefined UEs (pUEs) with stochastic predefined schedules and an intelligent UE (iUE) with an undefined transmission schedule, all transmitting over a shared, time-varying uplink channel. Jammers transmit unwanted packets to disturb the pUEs’ and the iUE’s communication. The iUE’s learning process is based on the deep reinforcement learning (DRL) framework, utilizing a residual network (ResNet)-based deep Q-Network (DQN). To coexist in the network and maximize the network’s sum cross-layer achievable rate (SCLAR), the iUE must learn the unknown network dynamics while concurrently adapting to dynamic channel conditions.  
Our simulation results reveal that, with properly defined state space, action space, and rewards in DRL, the iUE can effectively coexist in the network, maximizing channel utilization and the network’s SCLAR by judiciously selecting transmission time slots and thus avoiding collisions and jamming.     
\end{abstract}

\begin{IEEEkeywords}
Deep reinforcement learning, medium access control, jamming attacks, residual neural network.
\end{IEEEkeywords}

\section{Introduction}
With the proliferation of wireless communication technologies and the growing demand for high-speed data transmission, modern wireless networks have become increasingly complex and diverse. They often consist of a wide range of devices and technologies, each with unique characteristics and requirements. As a result, effective network management has become essential to ensure that these networks operate efficiently and reliably. The medium access control (MAC) layer plays an important role in ensuring data is transmitted efficiently and reliably by managing channel access across the network. Channel access represents a fundamental challenge in modern wireless networks as it affects communication network performance and reliability. Channel access strategies aim to allocate the limited available spectrum among multiple competing user equipment (UEs) efficiently and fairly. However, channel access management becomes more difficult in the presence of malicious entities, such as jammers, that try to disrupt legitimate UEs' communications by jamming or interfering with their transmissions \cite{pirayesh2022jamming}. Jammers, in particular, can severely degrade channel utilization and network performance, e.g., throughput, delay, and quality of service (QoS) \cite{khadr2022jamming,9486955}.

Traditional channel access mechanisms such as time division multiple access (TDMA) and carrier sensing multiple access (CSMA) demonstrate poor performance in heterogeneous networks, because they cannot adapt to changing network conditions \cite{wu2022satmac,guo2022multi}. More specifically, such channel access mechanisms cannot coexist with other channel access protocols in heterogeneous networks, as they are incompatible with those protocols, which causes synchronization issues and results in network performance degradation. Along with TDMA and CSMA, newer communication schemes such as orthogonal frequency division multiplexing (OFDM) and its multiple access counterpart, orthogonal frequency division multiple access (OFDMA), also face challenges in heterogeneous settings. Despite its advantages in terms of spectral efficiency, OFDMA requires precise synchronization and is sensitive to frequency offsets and phase noise. This can lead to interference and degrade network performance \cite{liu2021fine}. Furthermore, OFDMA does not possess inherent security features and is susceptible to malicious jamming \cite{yuan2023joint}.

To address these challenges, in this study, we propose using one intelligent user equipment (iUE) per cell that can learn to coexist with multiple UEs with predefined transmission schedules, referred to as predefined user equipment (pUEs), in the presence of jammers, and to implement a robust channel access strategy. The iUE use techniques such as machine learning (ML) and artificial intelligence (AI) to adapt to the dynamic and uncertain wireless environment and make smart decisions without requiring prior knowledge or explicit coordination. By learning the transmission schedules of pUEs and jammers' jamming behavior under constantly changing channel conditions, the iUE can implement an optimal transmission policy that maximizes channel utilization and the networks' sum cross-layer achievable rate (SCLAR). This approach addresses the concern of the network's performance degradation in traditional channel access mechanisms due to collisions, as well as provides a robust defense against malicious jamming.

In this paper, we closely analyze the implementation of such an iUE. We propose a deep reinforcement learning (DRL)-based iUE that can coexist with pUEs and random jammers in a multi-cell quasi-static wireless network. We refer to the considered network as ``quasi-static" due to its relatively stable physical and geographical configuration. However, the operational environment within this quasi-static network is dynamic and uncertain. In the considered network, the pUEs and the iUE transmit packets to a cluster head (CH) through a time-slotted uplink data channel, while the jammers try to disturb the UEs' transmissions to the CH. The data channel undergoes frequent transformations, meaning that all UEs' and jammers' transmit power, channel coefficients, and path loss parameters frequently vary. Consequently, each UE's signal-to-interference-plus-noise ratio (SINR) and achievable rate also change. On top of that, there are also intra-cell interference and inter-cell interference. Accordingly, the iUE tries to learn transmission schedules of pUEs and jammers' jamming procedures under constantly changing network conditions and implement an optimal transmission policy to maximize channel utilization and the network's SCLAR.
\subsection{Related Work}
Significant academic and industrial research efforts have been invested into developing ML-based solutions to address MAC-layer problems. For instance, \cite{miuccio2022learning} provided an overview of ML-based solutions for wireless networking applications. This study showed that radio resource management, particularly of the MAC parameters, significantly influences the performance of heterogeneous networks, and it advocates using ML-based approaches to improve network performance.

In \cite{da2020noma}, the authors proposed a Q-learning-based random access strategy incorporating non-orthogonal multiple access (NOMA) for ultra-dense networks. The solution involves a centralized base station (BS) that observes channel status to choose the optimal transmission power to access a given time slot.
Similarly, the authors of \cite{mennes2020multi} proposed a DRL-based solution to predict the spectrum occupancy of unknown neighboring networks in order to reduce inter-network collisions and improve the performance of cognitive radio networks (CRN). 

Furthermore, in \cite{xin2022deep}, the authors proposed a deep learning-based channel access and rate adaptation protocol for CSMA-based WiFi networks, while the authors of \cite{fihri2020machine} focused on using a supervised learning approach in CRNs to detect malicious attacks on the MAC layer that corrupt the contention mechanism. In the latter study, the authors used a support vector machine (SVM) to predict and identify an impending attack. However, while considering the malicious attacks in the channel access problem, the aforementioned study did not take into account varying channel conditions or experience-based learning to improve network performance.

Furthermore, in \cite{yu2019deep}, the authors proposed a DRL-based adaptive multiple access protocol known as DLMA (DRL Multiple Access), where an intelligent node can harmoniously coexist in a network with slot-based protocols like S-Aloha and TDMA by observing the environment, its actions, and the rewards that arise from interactions. However, in the aforementioned study, the primary focus was on legitimate UEs and jammers or time-varying channel characteristics were not accounted for.

In addition, several other notable works used DRL for anti-jamming in wireless communications. For instance, Xu et~al. \cite{xu2020intelligent} proposed an intelligent anti-jamming scheme for cognitive radio networks based on DRL. The authors designed a Double Deep Q Network (Double DQN) to model the confrontation between the cognitive radio network and the jammer, demonstrating effective defense against jamming attacks.

Likewise, Ye et~al. \cite{ye2020t} proposed a ($\tau, \epsilon$)-greedy reinforcement learning algorithm for anti-jamming wireless communications. Their approach, which adjusts the probability of choosing the previous action based on its value, demonstrated a faster convergence rate and a slightly higher signal-to-interference-plus-noise ratio as compared to $\epsilon$-greedy when applied to various DQN variants.

Furthermore, Ali et~al. \cite{ali2023defeating} proposed a DRL-based approach to defeat proactive jammers in resource-constrained internet of things (IoT) networks. To this end, aiming to identify the most robust and the least complex agent suitable for power-constrained devices, the authors trained different variants of DQN agents to mitigate the effects of jamming. The results of using this approach demonstrated effectiveness against proactive jammers, regardless of their jamming strategy. However, in the present study, we introduce additional complexities by considering a multi-cell scenario with multiple jammers, and specifically address both intra-cell and inter-cell interference.

In our previous work \cite{basit2024drl}, we proposed a DRL-based solution for channel access under jamming in a single-cell scenario. The present study extends our solution to a multi-cell scenario with multiple jammers, accounting for both intra-cell and inter-cell interference. In addition, in the present study, we incorporate the analysis of action selection from the learned policy and the theoretical structure of the proposed DRL framework. These advancements provide a more realistic and complex network environment with dynamic channel variations, thereby solidifying the substantial contribution of this work.

While the studies briefly reviewed above showcase diverse applications of DRL in MAC layer problems, most prior works focus on single-cell scenarios with limited users and jammers, thus frequently overlooking the complexities of dynamic channel characteristics. This being said, previous studies, along with our current research, underscore the growing interest and significant advancements in the application of DRL for channel access and anti-jamming in dynamic channel conditions. However, our work further extends these studies by considering a multi-cell scenario with multiple jammers and by accounting for both intra-cell and inter-cell interference, thereby contributing to the substantial body of knowledge in this field.

\subsection{Research Question and Problem of Interest}

Many approaches have been used to study the problem of accessing a shared medium with finite resources. However, most of them suffer from significant drawbacks, which compels us to explore advanced RL techniques as potential solutions to the medium access management problem. 

The inherent problem with using optimization-based algorithms is oversimplifying the dynamic environment by making several assumptions that do not model the dynamic nature of heterogeneous wireless networks. Mathematical optimization-based methods are generally state-oblivious. They, therefore, cannot model all the factors that impact the diverse and dynamic wireless environment with a single simplified model. Supervised learning-based methods, for their part, need a large amount of appropriately labeled data to be used for medium access management. However, in modern networks, acquiring such data is nearly impossible due to its highly dynamic and heterogeneous nature.

We address these challenges by proposing a DRL-based algorithm that enables the iUE to coexist harmoniously with the pUEs and jammers with the aim of enhancing channel utilization and the network's SCLAR by minimizing collisions and jamming susceptibility. Our methodology involves formulating the channel access problem as a partially observed Markov decision process (POMDP) under constantly changing channel conditions with jammers, employing a residual neural network (ResNet)-based deep Q-network (DQN) for optimal action-value function approximation, and conducting simulations to demonstrate that the proposed approach outperforms the benchmark solutions and achieves near-optimal performance by effectively reducing the collisions and avoiding jamming.
\subsection{Contributions}
This paper's major contributions can be summarized as follows:
\begin{itemize}
    \item We propose a novel DRL-based channel access mechanism for multi-cell quasi-static wireless networks in the presence of intra-cell and inter-cell interference, dynamic channel conditions, and malicious jamming. This mechanism addresses the challenging scenario of an iUE coexisting with legitimate pUEs and jammers. Using DRL, the iUE gains the ability to learn and adapt its transmission strategies through interactions with pUEs and jammers, as well as via identifying the patterns of the dynamic channel. 

    \item We formulate the problem of maximizing the network's SCLAR in the conditions of partial observability as a POMDP. This formulation accounts for the dynamic channel conditions, interference, and unknown transmission schedules of the coexisting pUEs and jammers. 

    \item To solve the formulated POMDP, we propose a novel ResNet-based DQN architecture specifically tailored to the problem of SCLAR maximization in multi-cell wireless networks with jammers. This architecture leverages the capabilities of ResNet to effectively extract features from the partially observed environment, including dynamic channel, interference and unknown transmission actions of coexisting entities. 

    \item The effectiveness of the proposed approach is validated through extensive simulations. The results demonstrate near-optimal performance in terms of channel utilization, SCLAR, and jamming robustness. In addition, we analyze how varying network parameters, including frame sizes and the combinations of pUEs and jammers, influence the performance of the proposed approach. This analysis showcases our method's adaptability in diverse network scenarios and provides valuable insights for a practical network configuration.
\end{itemize}

The rest of this paper is organized as follows. Section \ref{section: system model} presents the system model. The optimization problem is formulated as POMDP in Section \ref{section: problem_formulation}. Section IV proposes a residual deep Q-network (ResDQN)-based approach. The simulation results and performance evaluation are presented in Section \ref{section: simulation_results}, and the paper is concluded in Section \ref{section: conclusions}.

\textbf{Notations}: In this paper, vectors and matrices are represented by Boldface lowercase and uppercase letters, respectively. $|.|$ and $||.||$ stand for the absolute values of a scalar and the Euclidean norm of a vector or matrix, respectively. $\mathbb{E}[.]$ denotes the expectation operator. $\mathbb{C}^{M \times N}$ represents the space of complex-valued matrices whereas $\mathbb{R}^{M \times N}$ denotes the real-valued space. 
Table~\ref{table:notations} presents important notations and their description.    
\section{System Model}\label{section: system model}
\begin{table}[t]
\caption{Table of notations}
 \begin{tabular}{p{1cm} p{7cm}} 
\toprule \textbf{Notations} & \textbf{Descriptions}            \\ \hline
 $\mathrm{C}^{[k]}$  &  Cluster head of $k$-th cell \\  
 $N^{[k]}_{\tt UE}$  &   All legitimate UEs in $k$-th cell  \\  
 $M^{[k]}_{\tt J}$ & All jammers in $k$-th cell \\
 
 $\mathrm{UE}^{[k]}_n$    &  $n$-th UE in $k$-th cell             \\ 

 $\mathrm{J}^{[k]}_m$              &  $m$-th jammer in the $k$-th cell             \\ 
   $[\vec{A}^{[k]}_{\tt UE}]_{n,n}$ &  Transmission status of $\mathrm{UE}^{[k]}_n$             \\
 
 $\vec{A}^{[k]}_{\tt UE}$ & Cell operation matrix of all $N^{[k]}_{\tt UE}$ UE's in $k$-th cell \\ 
 $\vec{A}^{[i]}_{\tt UE}$ & Cell operation matrix of all $N^{[i]}_{\tt UE}$ UE's in $i$-th cell \\ 

  $[\vec{I}^{[k]}_{\tt J}]$              &  Jamming matrix of all $M^{[k]}_{\tt J}$ jammers in $k$-th cell             \\ 

 $[\vec{I}^{[k]}_{\tt {J}}]_{m,m}$              &  Jamming status of  $\mathrm{J}^{[k]}_m$ in $k$-th cell             \\

  $[\vec{I}^{[i]}_{\tt {J}}]_{m,m}$              & Jamming status of  $\mathrm{J}^{[k]}_m$ in $i$-th cell              \\

 $\vec{h}^{[k]}_{\mathtt{UE}_n}$              & Channel vector of $\mathrm{UE}^{[k]}_n$ $\rightarrow$ $\mathrm{C}^{[k]}$ in $k$-th cell             \\ 

 $\vec{h}_m^{[k]}$              & Channel vector of $\mathrm{UE}^{[k]}_m$ $\rightarrow$ $\mathrm{C}^{[k]}$ $\forall m \neq n$ in $k$-th cell             \\ 

 $\vec{h}_n^{[i]}$              & Channel vector of $\mathrm{UE}^{[i]}_n$ $\rightarrow$ $\mathrm{C}^{[i]}$ in $i$-th cell              \\ 

 $\vec{h}_m^{[i]}$              & Channel vector of $\mathrm{UE}^{[i]}_m$ $\rightarrow$ $\mathrm{C}^{[i]}$  $\forall m \neq n$ in $i$-th cell            \\ 
 
  $\vec{H}_{\mathtt{UE}}^{[k]}$             &  Channel matrix of all $N^{[k]}_{\tt UE}$ UE's $\rightarrow \mathrm{C}^{[k]}$               \\ 

  $\vec{H}_{\mathtt{UE}}^{[i]}$             &  Channel matrix of all $N^{[i]}_{\tt UE}$ UE's $\rightarrow \mathrm{C}^{[i]}$            \\ 

 $\vec{H}^{[k]}_{\tt{J}}$ & Channel matrix of all $M^{[k]}_{\tt J}$ jammers $\rightarrow \mathrm{C}^{[k]}$  \\ 

  $\vec{H}^{[i]}_{\tt{J}}$ & Channel matrix of all $M^{[k]}_{\tt J}$ jammers $\rightarrow \mathrm{C}^{[i]}$  \\ 

    $x^{[k]}_{\mathtt{UE}_n}$ & Transmitted signal of a single $\mathrm{UE}^{[k]}_n$ $\rightarrow\mathrm{C}^{[k]}$ \\ 

      $x^{[i]}_{\mathtt{UE}_n}$ & Transmitted signal of a single $\mathrm{UE}^{[i]}_n$ $\rightarrow\mathrm{C}^{[i]}$\\

      $x^{[k]}_{\tt {J}_m}$ & Jamming signal of a single $\mathrm{J}^{[k]}_m$ jammer in $k$-th cell \\ 

$x^{[i]}_{\tt {J}_m}$ & Jamming signal of a single $\mathrm{J}^{[i]}_m$ jammer in $i$-th cell \\
  
  $\vec{x}_{\mathtt{UE}}^{[k]}$              & Transmitted signal vector containing $x^{[k]}_n$ of each $\mathtt{UE}^{[k]}_n$             \\ 

   $\vec{x}_{\mathtt{UE}}^{[i]}$              &   Transmitted signal vector containing $x^{[i]}_n$ of each $\mathtt{UE}^{[i]}_n$            \\ 

$\vec{x}_{\tt J}^{[k]}$ & Jamming signal vector of all $M^{[k]}_{\tt J}$ jammers in $k$-th cell \\ 

$\vec{x}_{\tt J}^{[i]}$ & Jamming signal vector of all $M^{[i]}_{\tt J}$ jammers in $i$-th cell \\ 

   $\hat{\vec{x}}^{[k]}_\mathtt{UE}$              &  Estimated version of $\vec{x}^{[k]}_\mathtt{UE}$ at $\mathrm{C}^{[k]}$             \\  
 $\vec{P}^{[k]}_{\mathtt{UE}}$              &  Transmit power matrix of all $N^{[k]}_{\tt UE}$ in $k$-th cell              \\ 

 $\vec{P}^{[i]}_{\mathtt{UE}}$              &  Transmit power matrix of all $N^{[i]}_{\tt UE}$ in $i$-th cell             \\ 
 
 $\vec{P}^{[k]}_{\mathtt{J}}$ & Transmit power matrix of all $M^{[k]}_{\tt J}$ in $k$-th cell \\

$\vec{P}^{[i]}_{\mathtt{J}}$ & Transmit power matrix of all $M^{[i]}_{\tt J}$ in $i$-th cell  \\ 
 
 $P^{[k]}_{\mathtt{J}_m}$ & Transmit power of single $\mathrm{J}^{[k]}_m$ in $k$-th cell \\ 
 $P^{[i]}_{\mathtt{J}_m}$ & Transmit power of single $\mathrm{J}^{[i]}_m$ in $i$-th cell\\ 
  
  ${P}_{\mathtt{UE}_{n}}^{[k]}$             &   Transmit power of $\mathrm{UE}^{[k]}_n$ in $k$-th cell             \\ 
  ${P}_{\mathtt{UE}_{n}}^{[i]}$             &   Transmit power of $\mathrm{UE}^{[i]}_n$ in $i$-th cell            \\ 
  
 $\vec{y}^{[k]}$             &   Received signal vector at $\mathrm{C}^{[k]}$            \\ 
 
 $\vec{n}^{[k]}$              &    AWGN vector at $\mathrm C_k$ in $k$-th cell         \\

 $\vec{V}^{[k]}$             &    Decoding matrix of all $N^{[k]}_{\tt UE}$ in $k$-th cell          \\ 
$\vec{v}^{[k]}_{\mathtt{UE}_n}$               &   Decoding vector of $N^{[k]}_{\tt UE}$            \\
$\mathtt{SINR}^{[k]}_{\mathtt{UE}_n}$ & $\tt{SINR}$ of $\mathrm{UE}^{[k]}_n$ before applying MF-SIC\\

$\widehat{\mathtt{SINR}}^{[k]}_{\mathtt{UE}_n}$ & $\tt{SINR}$ of $\mathrm{UE}^{[k]}_n$ after applying MF-SIC\\

$C^{[k]}_{\mathtt{UE}_n}$ & Achievable rate of $\mathrm{UE}^{[k]}_n$ \\
$\vec{C}^{[k]}_{\mathtt{UE(sum)}}$ &Sum achievable rate of all $N^{[k]}_{\tt UE}$ UEs  \\

$
\xi^{[k]}_{\mathtt{UE}_n}[t^s_f]$ & Successful transmission rate of $\mathrm{UE}^{[k]}_n$ in time slot $t^s_f$\\

$R^{[k]}_{{\tt UE}_n}[t^s_f]$ & Cross-layer achievable rate of $\mathrm{UE}^{[k]}_n$ in time slot $t^s_f$\\
\bottomrule 
\end{tabular}
\label{table:notations}
\end{table}
\subsection{Uplink Multi-Cell Multi-User Multiple Input Single Output (UL MC-MU-MISO) Network}
\begin{figure}[!t]
\centering
 \includegraphics[width=\linewidth]{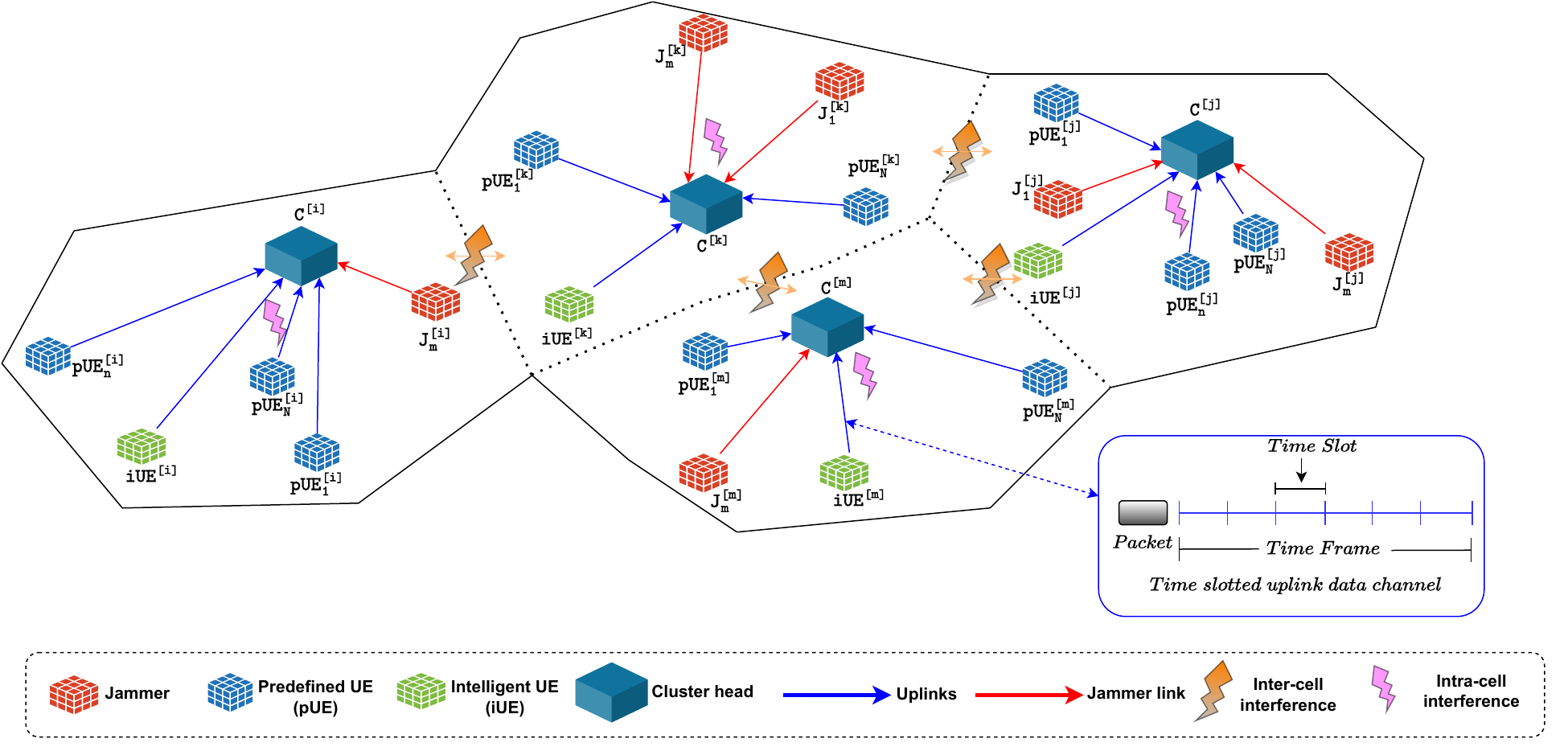}
\caption{Illustration of the considered network in the presence of malicious jammers.}
 \label{fig:network_model}
 \end{figure}
We consider a heterogeneous wireless network consisting of $K$ different cells. Let $\mathrm{C}^{[k]}$ be the CH of cell $k \; \forall \; k=\{1,2,\cdots, k, \cdots, K\}$. In cell $k$, there are \mbox{$N^{[k]}_{\tt UE}$ legitimate UEs} and \mbox{$M^{[k]}_{\tt J}$ jammers}. The legitimate UEs include some pUEs and an iUE. The iUE is considered to be saturated with data packets, i.e., it always has data packets available for transmission. The \mbox{$N^{[k]}_{\tt UE}$ legitimate UEs} transmit data packets to $\mathrm{C}^{[k]}$ via a shared time-slotted \mbox{($t= \{0, 1, 2, 3, \cdots\}$)} wireless uplink channel. In contrast, the $M^{[k]}_{\tt J}$ jammers persistently attempt to attack the data channel with the sole intention of disrupting communication between the legitimate UEs and $\mathrm{C}^{[k]}$, as illustrated in Fig.~\ref{fig:network_model}. Furthermore, we consider that there is not only intra-cell interference in the same cell but also inter-cell interference from other cells in the network. Let $\vec{A}^{[k]}_{\tt UE}$ denote the cell operation-action matrix of the $N^{[k]}_{\tt UE}$ UEs in the $k$-th cell. The $n$-th legitimate UE in the $k$-th cell is denoted by \mbox{$\mathrm{UE}^{[k]}_n \; \forall \; n=\{1, 2,\cdots,N^{[k]}_{\tt UE}\}$}. Let \mbox{$[\vec{A}^{[k]}_{\tt UE}]_{n,n}\in \{0, 1\}$} denote the transmission status of $\mathrm{UE}^{[k]}_n$, where $1$ means that the UE is transmitting a data packet in time slot $t$, and $0$ means that it is not. The $m$-th jammer in the $k$-th cell is denoted by \mbox{$\mathrm{J}^{[k]}_m \; \forall \; m=\{1, 2,\cdots,M^{[k]}_{\tt J}\}$.} The jamming status of  $\mathrm{J}^{[k]}_m$ can be denoted by \mbox{$[\vec{I}^{[k]}_{\tt J}]_{m,m}\in \{0, 1\}$}, where $1$ indicates the jammer is active, and $0$ indicates it is inactive. In each time slot, the $\mathrm{C}^{[k]}$ broadcasts an acknowledgment (ACK) to the UEs over a separate control channel. It is assumed that the jammers have no impact on the control channel. Next, we present the structure of the time frame and time slots. Then, we elaborate on the operating mechanism of the pUEs, the jammers, and the iUE. 
\subsection{Slotted Time-Based MAC}
We consider a discrete-time system in which time is divided into time slots of equal duration. The network operates on time synchronization, which allows the $\mathrm{UE}_n^{[k]}$ to transmit data packets to $\mathrm{C}^{[k]}$ in each time slot using a shared uplink data channel. 

The total time is split into \mbox{$T= \{t_1, t_2, \cdots, t_f, \cdots, t_F\}$} repeated frames, where the subscript \mbox{$\{1,2, \cdots, f, \cdots, F\}$} denotes the frame number and $F$ represents the total number of frames. Moreover, a single frame consists of a fixed number of time slots. For instance, the $f$-th frame is divided into \mbox{$t_f = \{t^1_f, t^2_f, \cdots, t^s_f, \cdots, t^S_f\}$} slots, where the superscript \mbox{$\{1,2, \cdots, s, \cdots, S\}$} represents the time slot number and $S$ denotes the total number of time slots in the frame, i.e., $t^s_f$ is the $s$-th time slot in the $f$-th frame. Additionally, the packet size of $\mathrm{UE}_n^{[k]}$ depends on the payload and other radio characteristics and is considered the same for all UEs. We assume that packet transmission starts at the beginning of a time slot, and a single packet takes up to one time slot to transmit. Furthermore, each UE can transmit multiple packets in a single frame. Thus, multiple UEs can share a radio frequency channel at any time slot $t^s_f$. In the event UEs transmit in different time slots, there is no packet collision, and each UE's data is received correctly at $\mathrm{C}^{[k]}$. However, in the event multiple UEs, i.e., \mbox{$\mathrm{UE}_n^{[k]}$ and $\mathrm{UE}_{n'}^{[k]} \in \mathcal{N} \backslash \{\mathrm{UE}_n^{[k]}\}$}, try to transmit simultaneously in a time slot, a packet collision occurs. Additionally, if, in a given time slot, \mbox{$[\vec{I}^{[k]}_{\tt J}]_{m,m}$ = 1} and \mbox{$\mathrm{UE}_n^{[k]}$} transmits a packet, jamming occurs and $\mathrm{UE}_n^{[k]}$'s packet is lost. Furthermore, we consider a quasi-static saturated network, which is detailed next.
\subsection{Quasi-Static Networking}
 We consider a quasi-static wireless network composed of pUEs, an iUE and jammers, where the pUEs follow the same transmission schedule for a set period of time. Such networks are commonly found in applications like smart grid systems, in which devices such as smart meters have relatively stable configurations and infrequent changes in transmission schedules. However, these networks often operate in challenging environments with dynamic channel conditions due to environmental factors and are susceptible to inter-cell and intra-cell interference, and malicious jamming. In the considered network, pUEs transmit according to their predefined schedules, representing devices with stable communication patterns over wide geographic areas. The iUE in the network uses our proposed channel access scheme to coexist with the pUEs and mitigate the impact of interference and jamming. The operations of both the legitimate UEs and the jammers are detailed below.
\subsubsection{Stochastic pUEs}
In the network in question, a pUE is a type of UE that has a predefined transmission schedule assigned to it in a stochastic manner. Consider a coefficient $[\chi]$ that equals 1 if a pUE transmits a packet in time slot $t^s_f$ and 0 otherwise. This coefficient models pUEs' transmission status, which follows a Bernoulli distribution that is parameterized by \mbox{$\Omega$, i.e., $[\chi] \sim $Bern$(\Omega)$.} Therefore, the signal to generate the $n$-th pUE's transmission vector in each time slot of a frame \mbox{$t_f= [0, 1, 0, 0, 1, ...]$} can be represented mathematically as
\begin{align} 
P([[\vec{A}^{[k]}_{\tt UE}]_{n,n}]^s_f) = 
    \begin{cases} 
        1 - \Omega &  [\chi] = 0 \\ \Omega & [\chi] = 1.
    \end{cases} 
\end{align}
\begin{figure}[t]
\includegraphics[width=\linewidth]{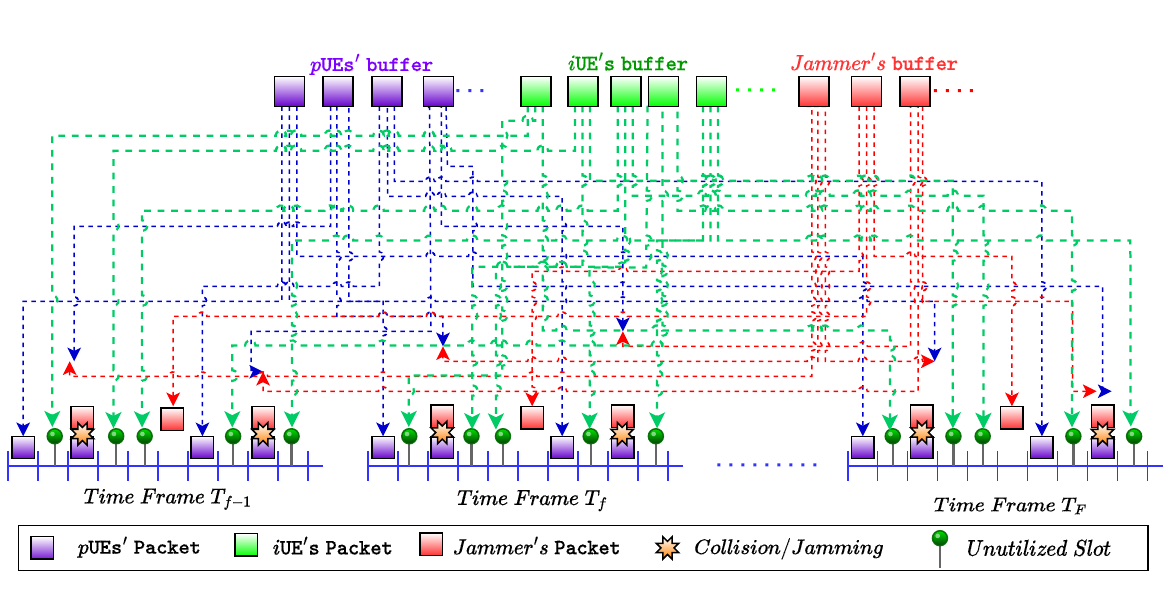}
\caption{Illustration of the time frame and transmission schedule of UEs and jammers in the considered network.}
\label{fig:time_frame_structure}
\end{figure}
\subsubsection{Random Jammers}
  We assume there are random jammers present in the network that transmit jamming signals intermittently according to a predetermined attack strategy with the sole intention of disrupting the pUEs' and the iUE's transmissions. Unlike a constant jammer, which transmits radio signals continuously, a random jammer alternates between jamming and sleeping. The jammer transmits in predefined $t_f^{s}$ time slots out of a total of $t^S_f$ slots and repeats its jamming pattern from frame to frame. Unlike a channel-aware jammer, a random jammer does not sense the network and only repetitively sends unauthenticated packets to disturb network transmission. Furthermore, each random jammer's status is independent of the other jammers in the network and transmission schedules of the legitimate UEs. The $\mathrm{J}^{[k]}_m$'s jamming status in slot $s$ of the $f$-th frame can be expressed as 
 \begin{align}
 [[\vec{I}^{[k]}_{\tt J}]_{m,m}]^s_f=\!\!
 \begin{cases}
 1, &\big[\!(n-1) S_{\tt{on}}\!\big] \!\!\leq\! t^s_f \!\leq\!\! \big[\!(n\!-\!1)\! S_{\tt{on}}\!\!+\!\!S_{\tt{off}}\big] \\ 
 0, &\big[(n-1) S_{\tt{on}} \!\!+\!\! S_{\tt{off}}\big] \leq t^s_f \leq [n S_{\tt{on}}],
 \end{cases}
 \end{align}
 where $n \in \mathbb{N}$ is the period number, which is set to a predefined value. Furthermore, \mbox{$S_{\tt{off}} \in S$}, \mbox{$S_{\tt{off}} < S_{\tt{on}}$}, and \mbox{$S_{\tt{off}} = [0, S_{\tt{off}}]$} is the time period in which the jammer is inactive and the transmission of \mbox{$\mathrm{UE}_n^{[k]}$} is thus unaffected. Similarly, \mbox{$S_{\tt{on}} \in S$}, and \mbox{$S_{\tt{on}} = [S_{\tt{on}}, S]$} is the time period in which the jammer is active. In the jammer's active period, the packets transmitted by \mbox{$\mathrm{UE}_n^{[k]}$} are destroyed. Moreover, the equation \mbox{$S_{\tt{on}} + S_{\tt{off}} = S$} holds true, which indicates that the sum of the jammer's active and inactive periods equals the total time period.
\subsubsection{DRL-Based iUE} 
The DRL-based iUE is a UE that uses our proposed channel access policy. The iUE is oblivious to 1) the number of coexisting pUEs, 2) the transmission schedules of the coexisting pUEs, 3) the presence and quantity of jammers, and 4) the jammers' operating procedure. The iUE aims to learn the coexisting pUEs' transmission schedules and the jammers' behavior to opportunistically implement an optimal transmission schedule in order to enhance the network's performance. 

Regarding channel access, the iUE differs from UEs that adopt contention-free MAC protocols because it does not have a predefined transmission schedule. It also differs from those that use contention-based protocols because it does not employ the back-off counter for re-transmission. Instead, it transmits its packets to $\mathrm{C}^{[k]}$ opportunistically, and in the event of collision or jamming, it re-transmits packets immediately without any wait time.  

The iUE generally learns from interactions with the network environment over a series of time frames. Then, it enacts a transmission strategy that enhances a given performance objective. Fig.~\ref{fig:time_frame_structure} shows an ideal (post-learning) scenario in which the iUE has successfully learned the optimal transmission schedule and utilized all unused time slots while completely avoiding jamming and collisions. 
\subsection{Signal Modelling in the Network}
We consider a multi-user system that consists of a CH, $\mathrm{C}^{[k]}$, and $N^{[k]}_{\tt UE}$ legitimate UEs (which include both the pUEs and the iUE) where \mbox{$N=\{1,2, \cdots, N^{[k]}_{\tt UE}\}$}. $\mathrm{C}^{[k]}$ is equipped with $L$
antennas, while each UE has a single antenna. We assume that all \mbox{$N^{[k]}_{\tt UE}$ UEs} share the same time-frequency resource. 
\subsubsection{Channel Modeling}
Let \mbox{$\vec{h}^{[k]}_{\mathtt{UE}_n} \in \mathbb{C}^{L \times 1}$} be the channel vector between the \mbox{$n$-th UE and $\mathrm{C}^{[k]}$.} Furthermore, the channel matrices between the \mbox{$N^{[k]}_{\tt UE}$ UEs} and \mbox{$\mathrm{C}^{[k]}$'s  $L$} antennas are written as \mbox{$\vec{H}_{\mathrm{UE}}^{[k]} \in \mathbb{C}^{L \times N^{[k]}_{\tt UE}}$.} In general, the propagation channel is modeled via large-scale and small-scale fading. However, we ignore the large-scale fading and assume that the elements of $\vec{H}^{[k]}_{\mathrm{UE}}$ are independent and identically Gaussian distributed with zero mean and unit variance. 

The path loss factor represents the attenuation or loss of signal due to distance or other factors such that $\vec{h}^{[k]}_{\mathtt{UE}_n}$ is proportional to $\frac{1}{d^2}$, where $d$ represents the distance between \mbox{$\mathrm{UE}^{[k]}_n$ and $\mathrm{C}^{[k]}$.} Fading is the channel coefficient or channel gain that represents the fading effect that is caused by the multi-path propagation of electromagnetic waves. In small-scale fading, $\vec{h}^{[k]}_{\mathtt{UE}_n}$ varies rapidly and this variation can be statistically described using a complex Gaussian distribution, which leads to $\vec{h}^{[k]}_{\mathtt{UE}_n}$ being Rayleigh distributed. Therefore, It is assumed that the amplitude of the received signal follows a Rayleigh distribution, which is suitable for scenarios in which there is no dominant line-of-sight path between the transmitter and the receiver \cite{jia2022rayleigh}. 
\subsubsection{Signal Received at $\mathrm{C}^{[k]}$}
\begin{figure*}
\begin{align} \label{eq_rx_signal}
 \vec{y}^{[k]}& = \vec{H}^{[k]}_{\mathtt{UE}} \left(\vec{P}^{[k]}_\mathtt{UE}\right)^{1/2} \vec{A}^{[k]}_{\tt UE}\vec{x}^{[k]}_{\mathtt{UE}} 
 + \vec{H}^{[k]}_{\mathtt{J}} \left(\vec{P}^{[k]}_{\mathtt{J}}\right)^{1/2}\vec{I}^{[k]}_{\mathtt{J}}\vec{x}^{[k]}_{\tt J}+ \sum_{\substack{i=1 \\i\neq k}}^{[k]}\vec{H}^{[i]}_\mathtt{UE} \left({\vec{P}_\mathtt{UE}^{[i]}}\right)^{1/2}\vec{A}^{[i]}_{\tt UE}\vec{x}_\mathtt{UE}^{[i]} + \sum_{\substack{i=1 \\ i\neq k}}^{[k]}\vec{H}^{[i]}_{\mathtt{J}} \left({\vec{P}^{[i]}_{\mathtt{J}}}\right)^{1/2}\vec{I}^{[i]}_{\mathtt{J}}\vec{x}_{\tt J}^{[i]}
 + \vec{n}^{[k]} \nonumber\\
 &= {\sum_{n=1}^{N^{[k]}_{\mathtt{UE}}} [\vec{A}^{[k]}_{\tt UE}]_{n,n} \left({P^{[k]}_{\mathtt{UE}_n}}\right)^{1/2} \vec{h}^{[k]}_{\mathtt{UE}_n} x^{[k]}_{\mathtt{UE}_n}}
 + \sum_{m=1}^{M^{[k]}_{\mathtt{J}}} [\vec{I}^{[k]}_{\tt J}]_{m,m} \left({P^{[k]}_{\tt {J}_m}}\right)^{1/2} \vec{h}^{[k]}_{\tt {J}_m} x^{[k]}_{\tt{J}_m} +\sum_{\substack{i=1, i\neq k}}^{[k]} \sum_{n=1}^{N^{[i]}_\mathtt{UE}} [\vec{A}_{\tt UE}^{[i]}]_{n,n} \left({P_{\mathtt{UE}_n}^{[i]}}\right)^{1/2} \vec{h}^{[i]}_{\tt UE_{n}} x^{[i]}_{\mathtt{UE}_n}
 \nonumber \\
 &\quad +\sum_{\substack{i=1, i\neq k}}^{[k]} \sum_{m=1}^{M^{[i]}_{\mathtt{J}}} [\vec{I}_{\tt J}^{[i]}]_{m,m} \left({P^{[i]}_{\tt{J}_m}}\right)^{1/2} \vec{h}^{[i]}_{\tt {J}_{m}} x^{[i]}_{\tt{J}_m}
 +\vec{n}^{[k]}.
\end{align}
\hrule
\end{figure*}
 \begin{figure*}
\begin{align} \label{eq_decoding_matrix}
	\hat{\vec{x}}^{[k]}_\mathtt{UE} &= (\vec{V}^{[k]})^{\sf H} \vec{y}^{[k]} =\textstyle (\vec{V}^{[k]})^{\sf H}\vec{H}^{[k]}_{\mathtt{UE}} \left({\vec{P}^{[k]}_\mathtt{UE}}\right)^{1/2} \vec{A}^{[k]}_{\tt UE}\vec{x}^{[k]}_{\mathtt{UE}} + (\vec{V}^{[k]})^{\sf H}\vec{H}^{[k]}_{\mathtt{J}} \left({\vec{P}^{[k]}_{\mathtt{J}}}\right)^{1/2}\vec{I}^{[k]}_{\mathtt{J}}\vec{x}_{\tt J}^{[k]} \nonumber\\ &\quad +\textstyle\sum_{\substack{i=1, i\neq k}}^{[k]}(\vec{V}^{[k]})^{\sf H} \vec{H}^{[i]}_\mathtt{UE} \left({\vec{P}_\mathtt{UE}^{[i]}}\right)^{1/2}\vec{A}^{[i]}_{\tt UE}\vec{x}_\mathtt{UE}^{[i]} + \textstyle\sum_{i=1,i\neq k}^{[k]}(\vec{V}^{[k]})^{\sf H}\vec{H}^{[i]}_{\mathtt{J}} \left({\vec{P}^{[i]}_{\mathtt{J}}}\right)^{1/2}\vec{I}^{[i]}_{\mathtt{J}}\vec{x}_{\tt J}^{[i]} + (\vec{V}^{[k]})^{\sf H} \vec{n}^{[k]}.
\end{align}
\hrule
 \end{figure*}
 \begin{figure*}
\begin{align} \label{eq_decode_signal}
 \hat{x}^{[k]}_{\mathtt{UE}_n}
 &= \underbrace{ [\vec{A}^{[k]}_{\tt UE}]_{n,n} \left({P^{[k]}_{\mathtt{UE}_n}}\right)^{1/2} \vec{v}^{[k]\sf H}_{\mathtt{UE}_n}\vec{h}^{[k]}_{\mathtt{UE}_n} x^{[k]}_{\mathtt{UE}_n}}_{\text{desired signal}}+\!\!\!\underbrace{\sum_{\substack{n'=1, n'\neq n}}^{N^{[k]}_{\tt UE}}\!\!\! [\vec{A}^{[k]}_{\tt UE}]_{n',n'} \left({P^{[k]}_{\mathtt{UE}_{n'}}}\right)^{1/2} \!\!\!\vec{v}^{[k]\sf H}_{\mathtt{UE}_n}\vec{h}^{[k]}_{\tt UE_{n'}} x^{[k]}_{\mathtt{UE}_{n'}}}_{\text{intra-cell interference}}
 +\!\!\! \underbrace{\sum_{m=1}^{M^{[k]}_{\mathtt{J}}} [\vec{I}^{[k]}_{\tt J}]_{m,m} \left({P^{[k]}_{\tt {J}_{m}}}\right)^{1/2}\!\!\!\vec{v}^{[k]\sf H}_{\mathtt{UE}_n} \vec{h}^{[k]}_{\tt {J}_m} x^{[k]}_{\tt{J}_{m}}}_{\text{intra-cell jamming}} \nonumber \\
 &\quad+ \underbrace{\sum_{\substack{i=1, i\neq k}}^{[k]} \sum_{n=1}^{N^{[i]}_\mathtt{UE}} [\vec{A}^{[i]}_{\tt UE}]_{n,n} \left({P^{[i]}_{\mathtt{UE}_n}}\right)^{1/2} \vec{v}^{[k]\sf H}_{\mathtt{UE}_n} \vec{h}^{[i]}_{\mathtt{UE}_{n}} x^{[i]}_{\mathtt{UE}_{n}}}_{\text{inter-cell interference}}
 +\underbrace{\sum_{\substack{i=1, i\neq k}}^{[k]} \sum_{m=1}^{M^{[i]}_{\mathtt{J}}} [\vec{I}^{[i]}_{\tt J}]_{m,m} \left({P^{[i]}_{\tt {J}_{m}}}\right)^{1/2} \vec{v}^{[k]\sf H}_{\mathtt{UE}_n}\vec{h}^{[i]}_{\tt {J}_{m}} x^{[i]}_{\tt {J}_m}}_{\text{inter-cell jamming}}
 + \underbrace{\vec{v}^{[k]\sf H}_{\mathtt{UE}_n}\vec{n}^{[k]}}_{\text{noise}}.
\end{align}
\hrule
\end{figure*}
\begin{figure*}
\begin{equation}\label{sinr_1}
  \mathtt{SINR}^{[k]}_{\mathtt{UE}_n} 
 = \frac{[\vec{A}^{[k]}_{\tt UE}]_{n,n} P^{[k]}_{\tt UE _n} |\vec{v}^{[k]\sf H}_{\mathtt{UE}_n} h^{[k]}_{\mathtt{UE}_n}|^2}
 { 
 \splitfrac{
 \bigg( 
 \sum\limits_{\substack{n'=1, n'\neq n }}^{N^{[k]}_{\tt UE}} [\vec{A}^{[k]}_{\tt UE}]_{n',n'} {P^{[k]}_{\tt UE_{n'}}} |\vec{v}^{[k]\sf H}_{\mathtt{UE}_n}\vec{h}^{[k]}_{\mathtt{UE}_{n'}}|^2 
 + \sum\limits_{\substack{j=1}}^{M^{[k]}_{\mathtt{J}}} \vec{I}^{[k]}_{\tt J_{m}} {P^{[k]}_{\tt {J}_{m}}}|\vec{v}^{[k]\sf H}_{\mathtt{UE}_n} \vec{h}^{[k]}_{\tt {J}_{m}}|^2 
 } 
 {
 + \sum\limits_{\substack{i=1,i\neq k}}^{[k]} \sum\limits_{\substack{n=1}}^{N^{[i]}_\mathtt{UE}} [\vec{A}_\mathtt{UE}^{[i]}]_{n,n} {P^{[i]}_{\tt UE_{n}}} |\vec{v}^{[k]\sf H}_{\mathtt{UE}_n} \vec{h}^{[i]}_{\tt UE_{n}}|^2 
 + \sum\limits_{\substack{i=1, i\neq k}}^{[k]} \sum\limits_{\substack{m=1}}^{M^{[i]}_{\mathtt{J}}} \vec{I}^{[i]}_{\tt J_{m}} {P^{[i]}_{\tt J_{m}}} |\vec{v}^{[k]\sf H}_{\mathtt{UE}_n}\vec{h}^{[i]}_{\tt J_{m}}|^2 + ||\vec{v}^{[k]}_{\mathtt{UE}_n}||^2 \delta^2 \bigg).}
 }
\end{equation}
\hrule
\end{figure*}
Let $x^{[k]}_{\tt UE_{n}}$ be the signal transmitted by $\mathrm{UE}^{[k]}_{n}$ to $\mathrm{C}^{[k]}$. Furthermore, the transmitted signal vector $\vec{x}_{\mathrm{UE}}^{[k]} \in \mathbb{C}^{N \times 1}$ is an $[N^{[k]}_{\tt UE} \times 1]$ column vector that contains the signals transmitted by each UE. 
The transmit power matrix \mbox{$\vec{P}^{[k]}_{\mathtt{UE}} = \operatorname{diag} ( P^{[k]}_1, \cdots, P^{[k]}_n, \cdots, P^{[k]}_N)$} is an \mbox{$[N^{[k]}_{\tt UE} \times N^{[k]}_{\tt UE} ]$} diagonal matrix that contains the transmit power of each UE in its diagonal elements, where ${P}_n^{[k]}$ represents the transmit power of the $n$-th UE in the $k$-th cell.

Since the $N^{[k]}_{\tt UE}$ UEs share the same time-frequency resource, the received signal vector $\vec{y}^{[k]}$ is an $[L \times 1]$ column vector that contains the signals received at each of $\mathrm{C}^{[k]}$'s $L$ antennas from all $N^{[k]}_{\tt UE}$ UEs. Therefore, the signal received at $\mathrm{C}^{[k]}$ is given by \eqref{eq_rx_signal}, where \mbox{$\vec{n}^{[k]} \in \mathbb{R}^{L \times 1}$}
denotes the additive white Gaussian noise (AWGN) vector at $\mathrm C^{[k]}$ where \mbox{$E[|\vec{n}\vec{n}^T|^2]=\sigma^2$}, $\vec{A}^{[k]}_{\tt UE}= \operatorname{diag} \big[[\vec{A}^{[k]}_{\tt UE}]_{1,1}, \cdots, [\vec{A}^{[k]}_{\tt UE}]_{{N^{[k]}_{\tt UE}},{N^{[k]}_{\tt UE}}}\big]$ is an $[N_{\tt UE}^{[k]} \times N_{\tt UE}^{[k]}]$ diagonal matrix, and 
\mbox{$\vec{I}^{[k]}_{\tt J}= \operatorname{diag} \big[[\vec{I}^{[k]}_{\tt J}]_{1,1}, \cdots, [\vec{I}^{[k]}_{\tt J}]_{M_{\tt J}^{[k]}, M_{\tt J}^{[k]}}\big]$} is an
$[M_{\tt J}^{[k]} \times M_{\tt J}^{[k]}]$
diagonal matrix.
\subsubsection{Linear Detection at $\mathrm{C}^{[k]}$ and SINR Formulation}
Consider \mbox{$\hat{\vec{x}}^{[k]}_\mathtt{UE} \in \mathbb{C}^{N^{[k]}_{\tt UE} \times 1}$} is estimated from $\vec{x}^{[k]}_\mathtt{UE}$ using a linear receiver. Also, considering the linear detection matrix \mbox{$\vec{V}^{[k]}\in \mathbb{C}^{L \times N^{[k]}_{\tt UE}}$} is used to separate the signal received at $\mathrm{C}^{[k]}$ into $N^{[k]}_{\tt UE}$ streams, the expression for $\hat{\vec{x}}^{[k]}_\mathtt{UE}$ can be given by~\eqref{eq_decoding_matrix}.

 The $n$-th column of matrix $\vec{V}^{[k]}$ represents the decoding vector of the $n$-th UE in the $k$-th cell, which is denoted as $\vec{v}^{[k]}_{\mathtt{UE}_n}$. Furthermore, $\vec{v}^{[k]}_{\mathtt{UE}_n}$ is used to detect the signal $x^{[k]}_{\mathtt{UE}_n}$, which we represent as the estimated signal $\hat{{x}}_{k}^\mathtt{UE}$ that is given by~\eqref{eq_decode_signal}. In our model, we consider two types of interference: inter-cell and intra-cell interference. Therefore, the SINR depends on the transmit power $\vec{P}$, the decoding vector $\vec{v}$, the channel vector $\vec{h}$, the overall interference, and the noise variance $\vec{n}$. The expression for SINR of the $n$-th UE in the $k$-th cell, $\mathtt{SINR}^{[k]}_n$, can be mathematically obtained as \eqref{sinr_1}.
\subsubsection{Matched Filter Successive Interference Cancellation for Signal Decoding at $\mathrm{C}^{[k]}$}
We employ matched filter successive interference cancellation (MF-SIC), which uses a matched filter as the decoding vector $\vec{v}^{[k]}_{\mathtt{UE}_n}$, to maximize the SINR of each UE by removing the interference from other UEs. More specifically, this technique cancels the interference from other UEs by subtracting their estimated signals from the received signal.
Thus, when matched filtering is used, the decoding vector for \mbox{$\mathrm{UE}_n^{[k]}$} can be obtained as
\begin{align} \label{mf_r1}
 \vec{v}^{[k]}_{\mathtt{UE}_n} &= \underset{{\vec{v}^{[k]}_{\mathtt{UE}_n}}}{\argmax} \frac{{P}^{[k]}_{\mathtt{UE}_n}|\vec{v}^{[k]\sf H}_{\mathtt{UE}_n} \vec{h}^{[k]}_{\mathtt{UE}_n}|^2}{||\vec{v}^{[k]}_{\mathtt{UE}_n}||^2 \delta ^2}. 
\end{align}
However, since 
\begin{align} \label{mf_r2}
\frac{P^{[k]}_{\mathtt{UE}_n}|\vec{v}^{[k]\sf H}_{\mathtt{UE}_n} \vec{h}^{[k]}_{\mathtt{UE}_n}|^2}{|\!|\vec{v}^{[k]}_{\mathtt{UE}_n}|\!|^2\sigma^2} \!\leq\! \frac{P^{[k]}_{\mathtt{UE}_n}|\!|\vec{v}^{[k]}_{\mathtt{UE}_n}|\!|^2 |\!|\vec{h}^{[k]}_{\mathtt{UE}_n}|\!|^2}{|\!|\vec{v}^{[k]}_{\mathtt{UE}_n}|\!|^2\sigma^2} \!\!=\!\! \frac{P^{[k]}_{\mathtt{UE}_n} |\!|\vec{h}^{[k]}_{\mathtt{UE}_n}|\!|^2}{\sigma^2},
\end{align}
therefore, we obtain the decoding vector with matched filtering from~\eqref{mf_r1} and~\eqref{mf_r2} as
\begin{align} \label{mf_r}
 \vec{v}^{[k]}_{\mathtt{UE}_n} = \beta^{[k]}_{\mathtt{UE}_n}||\vec{h}^{[k]}_{\mathtt{UE}_n}||^2,
\end{align}
where $\beta^{[k]}_{\mathtt{UE}_n}$ is a constant. 
\begin{figure*}
\begin{align}\label{mf_n}
  \widehat{\mathtt{SINR}}^{[k]}_{\mathtt{UE}_n}
 = \frac{[\vec{A}^{[k]}_{\tt UE}]_{n,n} P^{[k]}_{\tt UE_n} || h^{[k]}_{\mathtt{UE}_n}||^4} 
 {
\splitfrac{
 \bigg( \sum\limits_{\substack{n'=1\\n'\neq n }}^{N_{\tt UE}^{[k]}} [\vec{A}^{[k]}_{\tt UE}]_{n',n'} {P^{[k]}_{\tt UE_{n'}}} |\vec{h}^{[k]\sf H}_{\mathtt{UE}_n}\vec{h}^{[k]}_{\tt UE_{n'}}|^2 
 + \sum\limits_{\substack{m=1}}^{M^{[k]}_{\mathtt{J}}} \vec{I}^{[k]}_{\tt J_{m}} {P^{[k]}_{\tt J_{m}}}|\vec{h}^{[k]\sf H}_{\mathtt{UE}_n} \vec{h}^{[k]}_{\tt J_{m}}|^2}
 {+
 \sum\limits_{\substack{i=1\\i\neq k}}^{[k]} \sum\limits_{\substack{n=1}}^{N_i^\mathtt{UE}} [\vec{A}^{[i]}_{\mathtt{UE}}]_{n,n} {P^{[i]}_{\tt UE_{n}}} |\vec{h}^{[k]\sf H}_{\mathtt{UE}_n} \vec{h}^{[i]}_{\tt UE_{n}}|^2 
 + \sum\limits_{\substack{i=1\\i\neq k}}^{[k]} \sum\limits_{\substack{m=1}}^{M^{[i]}_{\mathtt{J}}} \vec{I}^{[i]}_{\tt J_{m}} {P^{[i]}_{\tt J_{m}}} |\vec{h}^{[k]\sf H}_{\mathtt{UE}_n} \vec{h}^{[i]}_{\tt J_{m}}|^2 + ||\vec{h}^{[k]}_{\mathtt{UE}_n}||^2 \delta^2\bigg)}}.
\end{align}

\hrule    
\end{figure*}
After matched filtering, $\mathrm{C}^{[k]}$ estimates the interference attributable to other UEs' signals. The estimated interference is then subtracted from the received signal to cancel it out. The interference cancellation process is performed successively for all interfering UEs starting with the one that interferes the most.
The received SINR of the $n$-th UE with MF-SIC can be obtained from~\eqref{sinr_1} and~\eqref{mf_r} as presented in \eqref{mf_n}.
\section{Problem Formulation}\label{section: problem_formulation}
\subsection{Physical-Layer Achievable Rate} A fundamental concept in communication theory is the physical-layer achievable rate. It is the highest data rate that can be transmitted over a channel with a specific modulation and coding scheme while maintaining a certain level of reliability or error rate \cite{perovic2021achievable}. The achievable rate considers the noise and interference in the channel as well as the specific modulation and coding scheme used in the communication system. The achievable rate generally shows how fast data can be transmitted with a certain quality. The achievable rate of $\mathrm{UE}^{[k]}_n$ in a single time slot $t^s_f$ can be represented mathematically as
\begin{align}\label{eq: physical_rate_slot}
 C^{[k]}_{\mathtt{UE}_n}[t^s_f] = \log_2 \left( 1 + \widehat{\mathtt{SINR}}^{[k]}_{\mathtt{UE}_n}[t^s_f] \right),
\end{align}
where $C^{[k]}_{\mathtt{UE}_n}[t^s_f]$ is the achievable rate in bits per slot per hertz. For the complete frame $t_f$, the achievable rate of a single $\mathtt{UE}^{[k]}_n$ can be expressed as
\begin{align}\label{eq: physical_rate_frame}
    \vec{C}^{[k]}_{\mathtt{UE}_n}[t_f]
    &=\textstyle \sum_{n=1}^{{N^{[k]}_{\tt UE}}} \sum_{s=1}^{S} C^{[k]}_{\mathtt{UE}_n}[t^s_f].
\end{align}
The sum rate of all $N^{[k]}_{\tt UE}$ UEs is a single frame, which includes the achievable rates of the individual $\mathtt{UE}^{[k]}_n$, is given by
\begin{align}
 \vec{C}^{[k]}_{\mathtt{UE(sum)}}[t^s_f] =\textstyle\sum_{n=1}^{N^{[k]}_{\mathtt{UE}}} C^{[k]}_{\mathtt{UE}_n}.
\end{align}

\subsection{Successful Transmission Rate}
The successful transmission rate ($\xi$) is a MAC-layer metric that measures the number of data packets that are successfully transmitted by \mbox{$\mathrm{UE}^{[k]}_n$ to $\mathrm{C}^{[k]}$} over a specified period of time. 
Let $p^{[k]}_{\mathtt{S}-\mathtt{UE}_n}[t^s_f]$ be the probability of successful transmission of $\mathtt{UE}^{[k]}_n \rightarrow \mathtt{C}^{[k]}$ in time slot $t^s_f$. Similarly, $p^{[k]}_{\tt I-\mathtt{UE}_n}[t^s_f]$ denotes the probability of idle transmission by $\mathtt{UE}^{[k]}_n$ in time slot $t^s_f$. Then $p^{[k]}_{\mathtt{S}-\mathtt{UE}_n}[t^s_f]$ depends on the idle probability of coexisting pUEs and the status of the jammers. Mathematically speaking, $p^{[k]}_{\mathtt{S}-\mathtt{UE}_n}[t^s_f]$ can be given by
\begin{align}\label{eq: success_prob}
p^{[k]}_{\mathtt{S}-\mathtt{UE}_n}[t^s_f] = g \big (p^{[k]}_{\tt {I}-\mathtt{UE}_1}, \cdots, p^{[k]}_{\tt {I}-\mathtt{UE}_n} , \vec{I}^{[k]}_{\tt J_1}\cdots \vec{I}^{[k]}_{\tt J_M} \big)[t^s_f].
\end{align}
On the other hand, the collision probability $p^{[k]}_{\tt C-\mathtt{UE}_n}$ in time slot $t^s_f$ can be computed as
\begin{align}\label{eq: coll_prob}
 p^{[k]}_{\tt C-\mathtt{UE}_n}[t^s_f] = 1 - p^{[k]}_{\tt I-\mathtt{UE}_n}[t^s_f] - p^{[k]}_{\mathtt{S}-\mathtt{UE}_n}[t^s_f].
\end{align}
By rearranging \eqref{eq: coll_prob}, the packet success rate $p^{[k]}_{\mathtt{S}-\mathtt{UE}_n}[t^s_f]$ can be expressed as
\begin{align}
 p^{[k]}_{\mathtt{S}-\mathtt{UE}_n}[t^s_f] = 1 - p^{[k]}_{\tt I-\mathtt{UE}_n}[t^s_f] - p^{[k]}_{\tt C-\mathtt{UE}_n}[t^s_f].
\end{align}
For the uplink transmission of $\mathrm{UE}^{[k]}_n \rightarrow \mathrm{C}^{[k]}$, $\xi$ can be calculated as the number of data packets successfully transmitted divided by the amount of time it took to transfer the packets. Thus, $\xi$ for $\mathrm{UE}^{[k]}_n$ in a single frame $t_f$ can be expressed as
\begin{align}
    \xi^{[k]}_{\mathtt{UE}_n}[t_f] =\frac{p^{[k]}_{\mathtt{S}-\mathtt{UE}_n}[t^1_f] + p^{[k]}_{\mathtt{S}-\mathtt{UE}_n}[t^2_f], \cdots, p^{[k]}_{\mathtt{S}-\mathtt{UE}_n}[t^S_f]}{S}.
\end{align}
\subsection{Cross-Layer Achievable Rate}
Cross-layer achievable rate (CLAR) is a term related to cross-layer wireless resource allocation, which is a technique that jointly considers the physical layer's achievable rate and the MAC layer's packet success rate in an integrated framework. The CLAR of $\mathtt{UE}^{[k]}_n$ in time slot $t^s_f$ is denoted by $R^{[k]}_{{\mathtt{UE}}_n}[t^s_f]$ and can be expressed as
 \begin{align}\label{eq_CLAR}
 R^{[k]}_{{\tt UE}_n}[t^s_f]= \xi^{[k]}_{\mathtt{UE}_n}[t^s_f] C^{[k]}_{{\tt UE}_n}[t^s_f].
\end{align}
 Furthermore, the CLAR for an entire frame, $\vec{r}^{[k]}_n \in \mathbb{R}^{T\times 1}$,  can be expressed as
\begin{align}
    \vec{r}^{[k]}_n = [R^{[k]}_n [t^1_f], R^{[k]}_n [t^2_f], \cdots, R^{[k]}_n [t^s_f] ,\cdots , R^{[k]}_n [t^S_f]].
\end{align}
 
 The network's SCLAR is the sum of the individual UEs' CLARs and depends on the transmission success and achievable rates of the pUEs and the iUE. In this work, we focus on solving the SCLAR maximization (SCLARM) problem, which we explain next.
\subsection{SCLARM Problem Formulation}
The SCLARM optimization problem for a given cell $k$ can be written as
\begin{subequations}\label{opt}
 \begin{alignat}{2}
 &\underset{\vec{A}^{[k]}_{\tt UE}}{\mathrm{maximize}}
 &\quad 
 & \sum_{n=1}^{|N^{[k]}_{\tt UE}|} (\vec{r}^{[k]}_n)^{\sf T} \vec{a}^{[k]}_{\tt{UE}_n},  \label{eq_obj_func}\\
 &\text{subject to} 
 && a^{[k]}_{\tt{UE}_n} (t^s_f) \in \{0, 1\}, \forall t^s_f \in t_f, \forall n \in N^{[k]}_{\tt {UE}}, \label{eq_constraint2} \\
 &&& \text{unknown } a^{[k]}_{\tt{J}_m} (t^s_f), \forall t^s_f \in t_f, \forall m \in M^{[k]}_{\tt {J}} \label{eq_constraint3}
 \end{alignat}
\end{subequations}
where $\vec{a}^{[k]}_{\tt{UE}_n} \!\!\!=\!\!\! \{\![\vec{A}_{\tt {UE}}^{[k]}]_{n} (T)\! \!:\!\!k \!\!=\!\!\! 1,..., \mathcal{C}, n\!\!\!=\!\!1,...,N^{[k]}_{\tt {UE}}, T \!\!\!=\!\! t_1,...,t_F\!\}$, represents the cell UE-action vector of $\mathrm{UE}^{[k]}_n$ over a given frame $t_f \in T$. In this context, $a^{[k]}_n(t^s_f)$ denotes the $\tt UE^{[k]}_n$'s action, and $a^{[k]}_{\tt{J}_m} (t^s_f)$ denotes the $J^{[k]}_m$'s action in 
a single slot $s$ belonging to that frame $t_f$. Lastly, $\vec{A}^{[k]}_{\tt UE}$, which corresponds to the transmission actions of all $N^{[k]}_{\tt UE}$ UEs is given by
    \begin{align} \label{eq_action_matrix}
 \vec{A}^{[k]}_{\tt{UE}} &= [\vec{a}^{[k]}_{\mathtt{UE}_1}, \cdots, \vec{a}^{[k]}_{\mathtt{UE}_n}, \cdots, \vec{a}^{[k]}_{\mathtt{iUE}} ] \nonumber\\& 
 =
 \begin{bmatrix}[lllll]
    a^{[k]}_{\mathtt{UE}_1} (t^1_f) & \dots  & a^{[k]}_{\mathtt{UE}_2} (t^1_f)  & \dots & a^{[k]}_{\mathtt{iUE}} (t^1_f) \\
    a^{[k]}_{\mathtt{UE}_1} (t^2_f) & \dots  & a^{[k]}_{\mathtt{UE}_2} (t^2_f)  & \dots & a^{[k]}_{\mathtt{iUE}} (t^2_f) \\
    \vdots & \dots  & \vdots  & \dots & \vdots \\
    a^{[k]}_{\mathtt{UE}_1} (t^S_f) & \dots  & a^{[k]}_{\mathtt{UE}_2} (t^S_f)  & \dots & a^{[k]}_{\mathtt{iUE}} (t^S_f) \\
 \end{bmatrix}.
\end{align}
\section{The POMDP Formulated and the DQN Solution Proposed}
In modern wireless networks, where changes are uncertain and stochastic, decision-making problems are typically modeled using the Markov Decision Problem (MDP) \cite{bozkus2023link}. This framework is particularly useful in dynamic environments, where conditions can change rapidly and unpredictably. A variant of MDP, partially observable MDP (POMDP), is used to model decision-making in environments in which the learning agent has incomplete information about the system's state. To be able to apply reinforcement learning (RL) algorithms to the channel allocation optimization problem, the problem, including the agent, the action space $\mathcal{A}$, the state space $\mathcal{S}$, the instantaneous reward function $r$, the transition probability $\mathcal{P}$, and the policy $\pi$, must be transformed to fit the POMDP framework.  

At time $t$, the agent observes a state $s_t \in \mathcal{S}$. The state space includes useful information about the wireless network environment, such as available radio resources, the number of UEs, their SINRs, and their transmission rates. Then, the agent takes action $a_t \in \mathcal{A}$. The action chosen must achieve network utility goals, such as spectral efficiency maximization or delay minimization. Then, the wireless environment transitions to a new state $s_{t+1}$ with a probability $\mathcal{P}$, and the agent receives an instantaneous reward $r_{t+1}$, which quantifies the quality of the action taken. The interaction ($s_t, a_t, r_{t+1}, s_{t+1}$) between the agent and the wireless environment continues, and the agent utilizes the instantaneous reward signal received to adjust its strategy until it learns the optimal policy $\pi^*$. The agent's policy $\pi$ defines the mapping from the states to the corresponding action $\mathcal{S} \rightarrow \mathcal{A}$, i.e., $a_t = \pi(s_t)$. Typically, the long-term reward is defined as the expected accumulated discounted instantaneous reward that is given by $\mathcal{R} = \mathbb{E}[\sum_{t=1}^{T} \gamma r_{t+1}(s_t, \pi(s_t))]$. The parameter $0 \leq \gamma \leq 1$ is the discount factor, which trades off instantaneous and future rewards. The agent's main goal is to obtain the optimal decision policy $\pi^*$ (i.e., to select the optimal radio resource) that maximizes the long-term reward, i.e., \mbox{$\pi^* = \max_\pi \mathcal{R}$.} 
 
Since the iUE (agent) in our SCLARM problem can see only the channel's current state at the beginning of each time slot, the full state of all time slots in the frame is not observable. However, the iUE can infer a distribution over the system state based on its past actions and observations. The SCLARM problem is, therefore, compatible with the general POMDP framework \cite{wang2018deep}. Next, we define the SCLARM problem's main parameters in terms of the POMDP framework. 

\subsubsection{Agent}

The agent is the iUE that uses our proposed algorithm to choose the appropriate transmission actions to maximize its objective function. The iUE's optimal strategy is to utilize the time slots that are not used by the coexisting pUEs while avoiding jamming. In time slot $t$, the iUE determines whether to transmit over the channel or wait depending on the observations it receives from $\mathrm{C}^{[k]}$. 
\subsubsection{Action Space}
A discrete action space $\mathcal{A} \dot=\{a_t: i = 1, 2\}$ represents the transmission choices that are available to the iUE at each time slot. We denote the action chosen by the iUE at time slot $t$ as $a_t \in \{\tt{dispatch}, \tt{hold} \}$, where $\mathtt{dispatch}$ means that the iUE transmits its packet in the time slot, and $\mathtt{hold}$ indicates that it does not. 
\subsubsection{Channel Observation}
 At time slot $t$, once the iUE has chosen an action $a_t \in \mathcal{A}$, $\mathrm{C}^{[k]}$ broadcasts ACK at $t+1$, denoted as $\mathrm{ACK}_{t+1}$, over a separate control channel. Specifically, $\mathrm{ACK}_{t+1}$ represents the iUE's transmission status (successful transmission, collision, or jammed transmission). Let us define the channel observation vector $\vec{c}_{t+1}$ to be an action-observation pair that can be expressed mathematically as
 \begin{align}
 \vec{c}_{t+1} = (a_{t} , \mathrm{ACK}_{t+1}).
 \end{align}
 Let $\vec{O}_{t+1}$ be the one hot representation of $\mathrm{ACK}_{t+1}$. Depending on the action $a_t$ that is chosen by the iUE, its unique possibilities are presented in Table \ref{table: channel state}.
\begin{table}[htp!]
\centering
\caption{Channel Observation Scenarios}
\label{table: channel state}
\begin{tabular}{lll}
\toprule
$a_t$ & $\mathrm{ACK}_{t+1}$ & $\vec{O}_{t+1}$ \\
\midrule
$\mathtt{dispatch}$ & Successful (S) & [$\mathtt{0, 0, 0, 0, 0, 1}$] \\ 
$\mathtt{dispatch}$ & Collision (C) & [$\mathtt{0, 0, 0, 0, 1, 0}$] \\ 
$\mathtt{dispatch}$ & iUE jammed ($J_A$) & [$\mathtt{0, 0, 0, 1, 0, 0}$] \\ 
$\mathtt{hold}$ & Idle (I) & [$\mathtt{0, 0, 1, 0, 0, 0}$] \\
$\mathtt{hold}$ & Busy (B) & [$\mathtt{0, 1, 0, 0, 0, 0}$] \\ 
$\mathtt{hold}$ & pUE jammed ($J_T$) & [$\mathtt{1, 0, 0, 0, 0, 0}$] \\
\bottomrule
\end{tabular}
\end{table}
\subsubsection{System State}
The state space \mbox{$\vec{\mathcal{S}} = [\vec{s}_1, \cdots, \vec{s}_t, \cdots, \vec{s}_T]$} contains all the possible underlying states that the environment can transition into; however it is not observable by the iUE. Since the environment is partially observable in our problem formulation, the iUE can see only a single system state $\vec{s}_t \in \vec{\mathcal{S}}$, which defines the environment's current status at time slot $t$. Partial observability is a challenge for the iUE since it has to infer the hidden information from the knowledge it has of previous actions and observations, i.e., $\vec{c}_{t-1}$. We initialize $\vec{s}_t$ with zero values, and, thus, the system state at time slot $t+1$, which is denoted by $\vec{s}_{t+1} \in \vec{\mathcal{S}}$, is given by 
\begin{align}\label{eq:environmental_state}
 \vec{s}_{t+1} &\!\!=\!\! \big[\!(\!\vec{c}_{t+1}^{\tt (UE_1)}, r_{t+1}^{\tt (UE_1)}\!), \!\cdots\!,\! (\!\vec{c}_{t+1}^{\tt (UE_n)} r_{t+1}^{\tt (UE_n)}\!) ,\!\cdots\!,\! (\!\vec{c}_{t+1}^{\tt (iUE)}, r_{t+1}^{\tt (iUE)}\!)\!\big],
\end{align}
where $\vec{c}_{t+1}^{\tt (UE_n)}$ corresponds to the action-observation pair of the $n$-th $\mathrm{UE}$, and $r_{t+1}^{\tt (UE_n)}$ is the reward received by the $n$-th $\mathrm{UE}$, which is explained later in this section. Note that $\vec{s}_{t+1}$ is the real-time representation of the iUE's transmission status and rate.
\subsubsection{Utility Functions}
We define the utility function of a pUE as being a function of the pUE's achievable rate. Mathematically speaking, the utility function of the $n$-th pUE at time slot $t$ is given by 
\begin{align}\label{eq: utility_legit}
\mathbb{U}_{t}^{\mathtt{UE}_{n}} = 
 \begin{cases}
 \nu^\mathtt{pUE} \times R^{\mathtt{UE}_n}, & \mathtt{ACK}_{t+1} = \mathrm{UE}_n \text{\;success} \\
 0 , & \mathtt{ACK}_{t+1} = \text{Otherwise},
 \end{cases}
\end{align}
where $\nu^\mathtt{pUE}$ is a scaling factor set out in Table~\ref{table: scaling_factors} and $R^{\mathtt{UE}_n}$ is the cross-layer achievable rate (CLAR) of the $n$-th $\mathrm{UE}$ that can be computed using \eqref{eq_CLAR}.  
Intuitively, \eqref{eq: utility_legit} implies that in the event a pUE successfully transmits a packet in a time slot, its utility is equal to its transmission rate at that time slot scaled by ${\nu^\mathtt{pUE}}$. 

However, the iUE's utility factors in a variety of scaling levels that influence the iUE's long-term objective. More specifically, we want the iUE to learn to avoid collisions with pUEs and evade jamming by choosing the actions that yield the highest utility. Mathematically speaking, the iUE's utility can be given by
\begin{align}
\mathbb{U}^{\tt iUE}_{t} = 
 \begin{cases}
  \nu^{\tt iUE} \times R^{\tt{iUE}}, & \mathtt{ACK}_{t+1} \in \{ \tt {E} \} \\
 {\nu^{\tt iUE}} \times R^{\tt{iUE}}, & \mathtt{ACK}_{t+1} \in \{ \tt{G}\} \\
 {\nu^{\tt{iUE}}} \times R^{\tt{iUE}}, & \mathtt{ACK}_{t+1} \in \{ \tt{B}\} \\
 {\nu^{\tt{iUE}}} \times R^{\tt{iUE}}, & \mathtt{ACK}_{t+1} \in \{ \tt{W} \}, \\
 \end{cases}
\end{align}
where ${\nu^{\tt \tt{iUE}}}$ is a scaling factor for the iUE's achievable rate that is given in Table \ref{table: scaling_factors}. More specifically, ${\nu^{\tt{iUE}}}$ enables the iUE to distinguish among the network's different scenarios. 

\begin{table}[t!]
\centering
\caption{Scaling factor $\nu^{\tt{pUE}}$, $\nu^{\tt{iUE}}$ \& $\nu^{\tt net}$ and \textnormal{iUE} decision encoding}
\label{table: scaling_factors}
\begin{tabular}{p{2.1cm}p{1.6cm}p{1.6cm}p{0.2cm}p{0.2cm}p{0.4cm}}
\toprule
Channel status ($\mathcal{C}$) & iUE action $a_t$ & iUE decision & $\nu^{\tt pUE}$ &$\nu^{\tt iUE}$ & $\nu^{\tt net}$ \\
\hline
Jammed & \tt{hold} & Good (\tt G) & 0 &4 & 5 \\
pUE transmitting & \tt{hold} & Good (\tt G) & 1&4 & 5 \\
Unused & \tt{hold} & Worst (\tt W) &0&1 & -10 \\
Jammed & \tt{dispatch} & Worst (\tt W) &0& 1 & -10 \\
pUE transmitting & \tt{dispatch} & Bad (\tt B) &0& 3 & -5 \\
Unused & \tt{dispatch} & Excellent (\tt E) &0& 5 & 10 \\
\bottomrule
\end{tabular}
\end{table}
\subsubsection{Reward Function}
The agent's reward function directly impacts the system objective \cite{ye2019deep}. Since the iUE's objective is to maximize the SCLAR, we associate the definition of reward with the sum of the utility values of all coexisting pUEs scaled by the scalar reward ${\nu^{\tt net}}$ that is defined in Table \ref{table: scaling_factors}. The value of reward $r_{t+1}$ at different values of $O_{t+1}$ is influenced by $\mathbb{U}^{\tt iUE}_{t}$ and $\sum_{n=1}^{N}\mathbb{U}_{t}^{\mathtt{UE}_{n}}$ such that
\begin{align}\label{eq: reward_function}
 r_{t+1} = 
 \begin{cases}
 \nu^{\tt net} \big(\mathbb{U}^{\tt iUE}_{t} + \overset{N}{\underset{n=1}{\sum}}\mathbb{U}_{t}^{\mathtt{UE}_{n}}\big) , & \mathtt{ACK}_{t+1} \in \{ \tt{E} \} \\
 {\nu^{\tt net}} \big(\mathbb{U}^{\tt iUE}_{t} + \overset{N}{\underset{n=1}{\sum}}\mathbb{U}_{t}^{\mathtt{UE}_{n}}\big), & \mathtt{ACK}_{t+1} \in \{ \tt{B} \} \\
 {\nu^{\tt net}} \big(\mathbb{U}^{\tt iUE}_{t} + \overset{N}{\underset{n=1}{\sum}}\mathbb{U}_{t}^{\mathtt{UE}_{n}}\big), & \mathtt{ACK}_{t+1} \in \{ \tt{G}\} \\
 {\nu^{\tt net}} \big(\mathbb{U}^{\tt iUE}_{t} + \overset{N}{\underset{n=1}{\sum}}\mathbb{U}_{t}^{\mathtt{UE}_{n}}\big), & \mathtt{ACK}_{t+1} \in \{ \tt{W} \}.
 \end{cases}
\end{align}
The reward function in \eqref{eq: reward_function} enables the iUE to learn an optimal transmission strategy that maximizes SCLAR. More specifically, $r_{t+1}$ encourages the iUE to utilize the unused time slots for transmission as much as possible yet avoid collisions with packets transmitted by coexisting pUEs and evade jamming.
\subsubsection{Transition Probability}
Let $\mathcal{P}$ be the state transition probability function representing the probability of switching from a state $s_t$ to the next state $s_{t+1}$ after the agent chooses an action $a_t$. More specifically, $\mathcal{P}: \mathcal{S} \times A \rightarrow \Pi(\mathcal{S})$, where $\Pi(\mathcal{S})$ is a probability distribution over some finite set, represents how the state of an environment changes with uncertainty, with the observation function connecting the system output (observations) to the true state of the system. Mathematically speaking, $\mathcal{P}$ can be given by
\begin{align}
\mathcal{P} = Pr{[s_{t+1}|s= s_t, a= a_t]}.
\end{align}
In our proposed system, $\mathcal{P}$ depends on the UEs' and the jammers' actions and is expressed as 
\begin{align}\label{eq: transition_prob}
 &\mathcal{P}(\vec{s}_{t+1} | \vec{s}_t, a_t) = p(a_{\tt UE_1}|\vec{s}_t) \times p(a_{\tt UE_2}|\vec{s}_t) \times   \cdots \times p(a_{\tt iUE}|\vec{s}_t) \nonumber \\
 &\quad\times \cdots \times p(a_{\tt UE_N}|\vec{s}_t)  \times p(a_{\tt J_1}|\vec{s}_t \cdots p(a_{\tt J_M} \times). 
\end{align}
Nevertheless, calculating the transition probability in \eqref{eq: transition_prob}, a key step in solving the optimization problem formulated in \eqref{eq_obj_func}, poses a significant challenge because the iUE is unaware of the network dynamics, including pUEs' transmission actions and the jammers' behavior. However, as it is explained in \cite{chen2020overview}, the historical trajectories that are generated from the iUE's interactions with the environment can be utilized to search for an optimal policy. Thus, the iUE can learn a policy $\pi$ directly from its experience interacting with the network environment instead of learning $\mathcal{P}$. 

\subsubsection{Policy}
Policy $\pi$ denotes the decision-making strategy through which an RL agent chooses an action $a_t$ based on the previous set of observations $O_t$. The RL agent aims to choose actions that maximize the expected discounted reward $\mathbb{E}[\overset{T}{\underset{k=0}{\sum}} \gamma^{[k]} r(s_{t+k}, a_{t+k})]$, where $r(s_{t+k}, a_{t+k})$ is the immediate reward received at time slot $t+k$.

The policy function $\pi(s)$ maps each state $s_t$ to a probability of selecting each possible action $a_t$.
If the agent follows policy $\pi(s)$ at time $t$, then $\pi(a_t|s_t)$ is the probability that $a_t=a$ if $s_t=s$. Different RL methods specify how the agent's policy changes in line with its experience. In general, the agent uses $\pi(s)$ to decide which action it should perform when it finds itself in state $s$.

Since we defined the iUE's action to be \mbox{$a_t \in \{\mathtt{dispatch}, \mathtt{hold} \}$}, we can define a policy matrix $\vec{\pi}(s)$ that contains one row for each state and one column for each action as follows
\begin{align}\label{eq: policy_matrix1}
 \vec{\pi}(s) =
 \begin{bmatrix}[ll]
 \vec{\pi}_{[s_1, \tt dispatch]} & \vec{\pi}_{[s_1, \tt hold]} \\
 \vec{\pi}_{[s_2, \tt dispatch]} & \vec{\pi}_{[s_2, \tt hold]} \\
 \vdots & \vdots \\
 \vec{\pi}_{[s_t, \tt dispatch]} & \vec{\pi}_{[s_t, \tt hold]} \\
 \vdots & \vdots 
 \end{bmatrix},
\end{align}
where $\underset{a_t}{\sum}\vec{\pi}_{s_t,a_t}= 1$ for each state $s_t$. Each element $\vec{\pi}_{s,a}$ in \eqref{eq: policy_matrix1} denotes the probability that action $a_t$ is taken when the iUE is in state $s_t$.

\subsection{The Approach Proposed to Solve the POMDP Formulated}
Value-based methods are used to estimate the agent's value function. The value function is then utilized implicitly to obtain $\pi^*$. Two value functions are employed in the literature to measure the accumulated discounted rewards expected for being in a certain state or taking a certain action in a state based on policy $\pi$: the state-value function $V^{\pi}(s)$ and the action-value function $Q(s_t, a_t)$. More specifically, the function $V^{\pi}(s_t)$ suggests the value of being in a state $s_t$ when following a policy $\pi$. Mathematically speaking, the state-value function $V^{\pi}(s)$, the action-value function $Q(s_t, a_t)$, the optimal value function $V^*(s)$, and the optimal action-value function $Q^*(s_t, a_t)$ can be expressed as
$V^{\pi}(s)\! =\! \mathbb{E}\big[\sum_{t=0}^{\infty}\!\gamma r_{t+1}(s_t, a_t, s_{t+1})|a_t \!\sim\! \pi(.|s_t), s_0\!=\!s\big],
    Q^{\pi}(s,a) =  \mathbb{E}\big[\sum_{t=0}^{\infty}\gamma r_{t+1}(s_t, a_t, s_{t+1})|a_t \sim \pi(.|s_t), s_0=s, a_0=a\big],
    V^*(s) = \max_{a_t}[r_{t+1}(s_t, a_t) + \gamma \mathbb{E}_{\pi}V^*(s_{t+1})],\, \text{and}\,
    Q^*(s,a) = r_{t+1}(s_t, a_t) + \gamma \mathbb{E}[\max_{a_{t+1}}Q^*(s_{t+1}, a_{t+1})]$
, respectively. 
In general, the RL agent's goal is to obtain the optimal policy $\pi^*(s) = \argmax_{a} Q^{*}(s, a)$. Various methodologies that each take their own unique approach can be employed to approximate $\pi^*(s)$. Value-based methods, for instance, determine $\pi^*(s)$ by updating $Q(s, a)$ until convergence is reached, where $Q(s, a) \approx Q^*(s, a)$ and subsequently derive $\pi^*(s)$ from $Q^*(s, a)$. A value-based method that is commonly used in wireless communication is Q-learning \cite{morales2020grokking}. We detail Q-learning and its extensions below.  
\subsubsection{Q-Learning}
In RL, Q-learning is one of the algorithms most widely used to solve problems modeled using the MDP. It is a model-free online RL algorithm that learns by interacting directly with the environment and updating its estimates based on the rewards and transitions it observes at each iteration. The entire learning procedure is based on figuring out the quality of each possible action and selecting an action accordingly. Thus, Q-learning tries to have complete and unbiased knowledge of all possible actions in all states. It therefore obtains the optimal Q-function values by iteratively using the following update rule that is based on the Bellman equation
\begin{align}\label{eq: q_learning_update_rule}
    Q(s_t, a_t) &\leftarrow Q(s_t, a_t) + \alpha[r_{t+1}(s_t, a_t) \nonumber \\
    &\qquad + \gamma\max_{a_{t+1}}Q(s_{t+1},a_{t+1}) - Q(s_t, a_t)],
\end{align}
where $\alpha$ is the learning rate that defines how much the new information contributes to the existing Q-value. The key concept behind the update rule in \eqref{eq: q_learning_update_rule} is finding the temporal difference (TD) between the current Q-value $Q(s_t, a_t)$ and the predicted Q-value $r_{t+1}(s_t, a_t) + \gamma\max_{a_{t+1}}Q(s_{t+1}, a_{t+1})$, which is the return expected after taking action $a_t$ in state $s_t$. The Q-learning algorithm uses the update rule in \eqref{eq: q_learning_update_rule} to construct a Q-table of all possible Q-values for each state-action pair. The algorithm is terminated when it reaches a certain number of iterations or when all Q-values converge. After convergence, the optimal policy determines the optimal action to take in each state, and $Q^{\pi^*}(s_t, a_t)$ is maximized for all states in $\mathcal{S}$, i.e., $\pi^* = \argmax_{a_{t+1}}Q^{\pi^*}(s_t, a_t)$. This optimal policy expresses that the action taken will eventually obtain the highest cumulative reward in any state.

However, traditional Q-learning cannot generalize well in dynamic environments and works only with low-dimensional state and action spaces \cite{yang2022distributed, yu2023user}. These limitations are overcome by utilizing deep Q networks (DQN) to approximate $Q(s, a)$ and employing techniques such as replay buffers and stochastic gradient descent.
\subsubsection{Deep Q Networks}

The deep Q network (DQN) algorithm was developed to overcome the limitations of tabular Q-learning and combines the advantages of both Q-learning and deep learning techniques \cite{sivaranjani2023artificial}. It seeks to replace the Q-table in Q-learning algorithms with a DQN that approximates $Q(s, a; \vec{\theta})$ by repeatedly adjusting the values of $\vec{\theta}$, which represents the neural network parameters, through training. 

The main idea behind DQN is to use function approximation to learn a parametric approximation $Q(s_t, a_t; \vec{\theta})$ of $Q(s_t, a_t)$. Function approximation makes it possible to approximately compute the action-value function for a given $\vec{\theta}$ for unseen state-action pairs $(s_t, a_t)$. The DQN learns $\vec{\theta}$ instead of the $|\mathcal{S}| \times |\mathcal{A}|$ matrix of Q-values. In other words, the neural network tries to find the values of $\vec{\theta}$ that approximate the Bellman equations for $ V^*(s)$ and $ Q^*(s, a)$ in every state-action pair $(s, a) \in \mathcal{S} \times \mathcal{A}$. We refer to the process of finding these values of $\vec{\theta}$ as the DQN training cycle and explain it next.  
\subsubsection{DQN Training Cycle}\label{sec: training_cycle}
 Assume that at time slot $t$, the state of the environment is $s_t$ and the DQN's training parameters are $\vec{\theta}$. The agent takes an action \mbox{$a_t = \argmax_a Q(s_t, a_t;\vec{\theta})$} for which the DQN gives an output of $Q(s_t, a_t;\vec{\theta})$ $\forall a_t \in$ $\mathcal{A}$. If the reward received by the agent is $r_{t+1}$ and the state transitions from $s_t$ to $s_{t+1}$, the tuple $(s_t, a_t, r_{t+1}, s_{t+1})$ forms an experience tuple that the DQN utilizes for training. The DQN finds the optimal values of $\vec{\theta}$ by minimizing the prediction errors of $Q(s_t, a_t; \vec{\theta})$ using a loss function. Mean squared error (MSE) is the most commonly used loss function \cite{patterson2022robust}. Therefore, for a specific experience tuple $(s_t, a_t, r_{t+1}, s_{t+1})$, loss $\mathcal{L}_{\tt MSE}(\vec{\theta})$ in the training process can be defined as
\begin{align}\label{eq:dqn_loss}
\mathcal{L}_{\tt MSE}(\vec{\theta}) &\!\!= \!\!\big(y^{\tt True} \!-\! y^{\tt Pred}\big)^2 \!\!=\!\!\big(y^{\tt True}_{r_{t+1},s_{t+1}} \!-\! Q(s_t, a_t ; \vec{\theta})\big)^2,
\end{align}
where $y^{\tt Pred} = Q(s_t, a_t;\vec{\theta})$ is the predicted Q-value for ($s_t, a_t$) given by DQN and $y^{\tt True}_{r_{t+1},s_{t+1}}$ is the DQN's target output, which can be calculated using the Bellman equation as follows 
\begin{align}\label{eq:target_output_DQN}
 y^{\tt True}_{r_{t+1},s_{t+1}} = r_{t+1} + \gamma \max_{a_{t+1}} Q(s_{t+1} , a_{t+1} ; \vec{\theta}). 
\end{align}
It should be noted that $(y^{\tt True}_{r_{t+1},s_{t+1}})$ is based on the current reward $r_{t+1}$ plus the discounted rewards predicted for the next observed states $\gamma max_{a_{t+1}} Q(s_{t+1} , a_{t+1} ; \vec{\theta})$ given by the environment. A semi-gradient algorithm is used to train the DQN by repeatedly updating $\vec{\theta}$. The iterative process of updating $\vec{\theta}$ can be mathematically expressed as
\begin{align}\label{eq:iterative_process_theta}
\vec{\theta}_{k+1} \leftarrow \vec{\theta}_k - \alpha_k \nabla_{\vec{\theta}} \mathcal{L}_{\tt MSE}(\vec{\theta}_k), 
\end{align}
where $\alpha_k$ is the step size at iteration $k$ and $\nabla_{\vec{\theta}} \mathcal{L}_{\tt MSE}(\vec{\theta}_k)$ is the gradient of the loss function with respect to $\vec{\theta}$ at iteration $k$. Note that $y^{\tt True}_{r_{t+1},s_{t+1}}$ in \eqref{eq:iterative_process_theta} is treated as a constant when computing the gradient because it is not a function of $\vec{\theta}$.

Nonetheless, as is mentioned in \cite{morales2020grokking}, updating $\vec{\theta}$ for a single experience can result in a less efficient learning process. Therefore, the experience replay technique and a quasi-static target network are used in DQN to ensure stability during the learning process, which leads to improved performance. Instead of training DQN with a single instance of experience replay, many instances are stacked together for batch training. An experience buffer ($\tt EB$) with a fixed storage capacity is used to store the experiences $e = (s_t, a_t, r_{t+1}, s_{t+1})$ gathered from different time slots in a first in first out (FIFO) manner, i.e., once the experience buffer is full, when the next experience is added, the oldest experience is removed from the buffer. For each training iteration, a mini-batch $\tt B$ consisting of $\tt N_{\tt BE}$ experiences is randomly sampled from the experience buffer to compute the loss function in \eqref{eq:dqn_loss} and update $\vec{\theta}$. 

Moreover, with the quasi-static target network technique, the DQN maintains two networks, namely a prediction network $Q(s_t, a_t, \vec{\theta}^{\tt pred})$ and a target network $Q(s_t, a_t; \vec{\theta}^{\tt target})$, where $\vec{\theta}^{\tt pred}$ and $\vec{\theta}^{\tt target}$ refer to the training parameters of their respective network. More specifically, the target network helps to stabilize the learning process by reducing the variance and correlation between the predicted Q-values. Thus, the DQN trains the prediction network at each learning step to decrease the loss function in \eqref{eq:dqn_loss}. The term $Q(s_t,a_t;\vec{\theta})$ for target output in \eqref{eq:dqn_loss} is computed using the target network, whereas the term $Q (s_{t+1}, a_{t+1}; \vec{\theta})$ in \eqref{eq:target_output_DQN} is computed using the prediction network.
 
The target network is typically a copy of the original DQN for which the training parameters $\vec{\theta}^{\tt target}$ get replaced by $\vec{\theta}^{\tt pred}$. However, to avoid the prediction network chasing a moving target network, we limit updating $\vec{\theta}^{\tt target} \leftarrow \vec{\theta}^{\tt pred}$
at each time slot. Instead of immediately following $\vec{\theta}^{\tt target}$ at each training iteration, the target network updates the values of $\vec{\theta}^{target}$ using a soft update rule \cite{kobayashi2021t}, which is expressed as   
\begin{align}\label{eq: update_theta_target}
    \vec{\theta}^{\tt target} \leftarrow (1-\tau)\vec{\theta}^{\tt target} + (\tau)\vec{\theta}^{\tt pred},
\end{align}
where $\tau$ is a hyper-parameter that controls the frequency of target updates. Using this technique results in $\vec{\theta}^{\tt target}$ being updated gradually and more smoothly after a few iterations. 
Using the experience replay and quasi-static target network techniques with the soft target update rule, we get
\begin{align}\label{eq: theta_update_final}
\vec{\theta}^{\tt pred}_{k+1} \leftarrow \vec{\theta}^{\tt pred}_k - \alpha_k \nabla_{\vec{\theta}} \mathcal{L}_{\tt MSE_{\tt B}}(\vec{\theta}^{\tt pred}_k; \vec{\theta}^{\tt target}_k),
\end{align}
\begin{align}\label{eq:target_with_e_replay}
 y^{\tt True}_{r_{t+1},s_{t+1}} = r_{t+1} + \gamma \max_{a_{t+1}} q (s_{t+1} , a_{t+1} ; \vec{\theta}^{\tt pred}), 
\end{align}
\begin{align}\label{eq:loss_with_e_replay}
    \mathcal{L}_{\tt MSE_{B}}(\vec{\theta}) = \frac{1}{\tt N_{\tt BE}}\sum_{e \in \tt EB}{\big(y^{\tt True}_{r_{t+1},s_{t+1}}-q(s_t,a_t;\vec{\theta}^{\tt target})}\big)^2.
\end{align}
In the next sections, we present the DQN and ResNet architectures used. 

\subsection{Neural Network Architecture}
Consider a k-layered neural network that represents a function $f(\vec{x}; \vec{\theta^{\tt pred}})$, where $\vec{x} \in \mathcal{D}_x \subseteq \mathbb{R}^{k_0}$ is the neural network's input, and $\vec{\theta^{\tt pred}} : \{\vec{W}, \vec{b}\}$ is a set of trainable parameters. Then, the neural network's output $\vec{y} \in \mathcal{D}_y \subseteq \mathbb{R}^{k_n}$ is given by $\vec{y} = f(\vec{x}; \vec{\theta^{\tt pred}})$, where $k_0$ represents the input dimension, and $k_n$ shows the output dimension. A neural network generally consists of an input layer, an output layer, and some hidden layers in between. Hence, a neural network can be represented by
\begin{align}
    f( \vec{x}; \vec{\theta^{\tt pred}}) = \big[ f_{\tt K} \circ f_{\tt K-1} \circ ... \circ f_{\tt k} \circ ... \circ f_{\tt 1}](\vec{x}; \vec{\theta^{\tt pred}}),
\end{align}
where the subscript $\tt K$ represents the last layer of the neural network and $\tt 1$ represents the input layer. 
\begin{Remark}
Since the loss function, which incorporates the objective function \eqref{eq_obj_func}, i.e., $\sum_{n=1}^{|\mathcal{N}^{[k]}|} (\vec{r}^{[k]}_n)^{\sf T} \vec{a}^{[k]}_n$, is not smooth w.r.t. the optimization variables $\vec{a}^{[k]}_n$, residual blocks help to smooth the loss function w.r.t. $\vec{a}^{[k]}_n$.
\end{Remark}
\subsubsection{Fully Connected Neural Network}
A fully connected neural network is a neural network in which every neuron in one layer is connected to every neuron in the next layer. Therefore, each input in the input vector $\vec{x}$ influences each output in the output vector $\vec{y}$. The output of each neuron is computed by applying a linear transformation followed by a non-linear activation function. Mathematically speaking, this operation can be expressed as
\begin{align}\label{eq: neuron_operation}
 \vec{a} = f_{\tt fc}(\vec{W}\vec{z} + \vec{b}),
\end{align}
where $\vec{z}$ is the input vector, $\vec{W}$ is the weight matrix, $\vec{b}$ is the bias vector, and $f_{\tt fc}$ is the activation function that adds some non-linearity so that the neural network can learn the complex model. Finally, $\vec{a}$ is the output vector, which can be the next layer's input. Thus, a fully connected neural network model with $k$ fully connected layers can be expressed as 
\begin{align}
    f( \vec{s}_t; \vec{\theta^{\tt pred}}) = \big[ f_{\tt fc_k} \circ f_{\tt fc_{k-1}} \circ ... \circ f_{\tt fc_1} ... ](\vec{\vec{s}_t}; \vec{\theta^{\tt pred}}),
\end{align}
where $\vec{s}_t$ is the input to the neural network and comes from \eqref{eq:environmental_state}, $\tt fc_k$ is the last fully connected layer, and $\tt fc_1$ is the first layer in the network.

The network's depth is generally one of many factors that can enhance neural network performance. Hence, we would like to analyze its impact on the iUE's learning performance, considering it interacts in real-time with the coexisting pUEs. Note, however, that deeper networks require more training time, experience more training loss and are less accurate \cite{sariturk2023comparison}, which is unsuitable for the system model we consider, in which the iUE must follow the stochastic transmission schedule of coexisting pUEs and jammers. Therefore, we leverage another powerful neural network, residual neural network (ResNet), that is able to overcome the degradation problem faced by deeper fully connected neural networks \cite{alaeddine2021deep}. Next, we present ResNet's architecture. 
\subsubsection{Residual Neural Network (ResNet)}
ResNet is another popular type of neural network that is used in many fields of application \cite{alaeddine2021deep} \cite{hu2021spiking}. In addition to having fully connected layers, a ResNet contains residual blocks. It uses skip connections to perform identity mapping between layers. A skip connection is a connection that bypasses one or more layers in a neural network and adds the input of the skipped layer(s) to the output of the next layer. As the number of residual blocks increases, the model's accuracy and generalization performance improve \cite{radosavovic2020designing}. 

Mathematically speaking, a ResNet with $L: \{l= 1, 2, \cdots, l, \cdots, L\}$ layers, \mbox{$R: \{r= 1, 2, \cdots, r, \cdots, R\}$} residual blocks, and \mbox{$M: \{m= 1, 2, \cdots, m, \cdots, M\}$} layers in each residual block can be expressed as
\begin{align}
    f(\vec{s_t}; \vec{\theta^{\tt pred}}) = [f_{\tt fc_L} \!\circ\! f_{\tt res_R} \!\circ\! \cdots \!\circ\! f_{\tt res1} \!\circ\! f_{\tt fc_1}](\vec{s_t}; \vec{\theta^{\tt pred}}),
\end{align}
where $f_{\tt fc_L}$ is the last fully connected layer, $f_{\tt res_R}$ is the last residual block, and $f_{\tt fc_1}$ is the first fully connected layer. Each residual block can be further decomposed into
\begin{align}
    f_{\tt res_r}(\vec{s_t}, \vec{\theta^{\tt pred}}) = f_{\tt fc_m}(\vec{s_t},\vec{\theta^{\tt pred}}) + f_{\tt res_{r-1}}(\vec{s_t},\vec{\theta^{\tt pred}}),
\end{align}
where $f_{\tt fc_m}$ is the last fully connected layer within the $r$-th residual block, and $f_{\tt res_{r-1}}$ is the previous residual block. The first residual block can be written as
\begin{align}\label{eq: residual_function}
    f_{\tt res_1}(\vec{s_t},\vec{\theta^{\tt pred}}) = f_{\tt fc_M}(\vec{s_t},\vec{\theta^{\tt pred}}) + \vec{s_t},
\end{align}
where $f_{\tt fc_M}$ is the last fully connected layer within the first residual block, and $\vec{x}$ is the ResNet's input vector. In addition, the output of each fully connected layer in a residual block can be calculated as follows
\begin{align}
    f_{\tt fc_m}(\vec{s_t},\vec{\theta^{\tt pred}}) = f_{\tt act_m}(f_{\tt dense_m}(\vec{s_t},\vec{\theta^{\tt pred}})),
\end{align}
where $f_{\tt dense_m}$ is a dense layer that applies a linear transformation to the input vector using \eqref{eq: neuron_operation}. Note that $f_{\tt dense_m}$ is equivalent to a fully connected layer, but we use it to refer to the last layer of a network that produces the final output.

The output of each fully connected layer that is outside of a residual block can be calculated as follows
\begin{align}
    f_{\tt fc_l}(\vec{s_t},\vec{\theta^{\tt pred}})) = f_{\tt act_l}(f_{\tt dense_l}(\vec{s_t},\vec{\theta^{\tt pred}}))),
\end{align}
where $f_{\tt act_l}$ is denotes the activation function at layer $l$.

For a given ResNet defined above, a skip connection in each residual block can be expressed as
\begin{align}
    f_{\mathtt{S}kip}(\vec{s_t},\vec{\theta^{\tt pred}}) = \vec{s_t} + f_{\tt res_r}(\vec{s_t},\vec{\theta^{\tt pred}}),
\end{align}
where $f_{\tt res_r}(\vec{s_t},\vec{\theta^{\tt pred}})$ is a residual function. The basic building block of a ResNet is a residual block, which consists of a few layers and a skip connection. The residual block learns the residual function $F(\vec{s_t},\vec{\theta^{\tt pred}}) = f_{\tt res_r}(\vec{s_t},\vec{\theta^{\tt pred}}) = \vec{s_t} + f_{\tt fc_m}(\vec{s_t},\vec{\theta^{\tt pred}})$ that represents the difference between the desired output $\vec{s_t} + f_{\tt fc_m}(\vec{s_t},\vec{\theta^{\tt pred}})$ and the input $\vec{s_t}$ instead of learning the direct mapping $\vec{s_t} + f_{\tt fc_m}(\vec{s_t},\vec{\theta^{\tt pred}})$. In other words, the residual block does not try to learn the exact output of the layer but rather the difference between the output and the input. The idea is that learning the residual function is easier than learning the direct mapping because the residual function can be zero or close to zero when the input is already close to the desired output. This enables the network to learn identity functions when needed, which makes it more flexible and robust. Furthermore, the residual function enables the gradient to flow directly from the output to the input, which improves training speed and network stability\cite{fang2021deep}. 
In general, a ResNet is constructed by stacking multiple residual blocks together. Thus, the cumulative residual function $f_{\tt res_R}(\vec{s_t},\vec{\theta^{\tt pred}})$ of $R$ residual blocks is obtained by applying the residual function of each block to the output of the previous block. It can be mathematically represented as
\begin{align}\label{eq: cummulative_residual_function}
    & f_{\tt res_R}(\vec{s_t},\vec{\theta^{\tt pred}}) \nonumber \\
    &=\sum_{r=1}^R f_{\tt fc_m}(\vec{s_t},\vec{\theta^{\tt pred}}) + \sum_{r=1}^{R-1} f_{\tt res_{R-r}}(f_{\tt fc_m}(\vec{s_t},\vec{\theta^{\tt pred}})),
\end{align}
where $f_{\tt fc_m}$ is the last fully connected layer in each residual block, and $f_{\tt res_r}$ is the function of the $r$-th residual block. More specifically, \eqref{eq: cummulative_residual_function} suggests that the cumulative residual function of $R$ blocks is obtained by adding the residual function of each block to the output of the previous block. The output of the ResNet can then be obtained by adding the input vector and the cumulative residual function.
\begin{align}
    f(\vec{s_t},\vec{\theta^{\tt pred}}) = \vec{s_t} + f_{\tt res_R}(\vec{s_t},\vec{\theta^{\tt pred}}).
\end{align}
\subsection{Proposed Algorithm} 
In our proposed solution, a ResNet-based DQN is used to approximate $Q(s_t, a_t; \vec{\theta^{\tt pred}}) \approx Q^*(s_t, a_t)$. The input to the neural network is a state $s_t$, and at the output, we obtain the approximated values of $Q = \{Q(s_t, a_t; \vec{\theta^{\tt pred}}) | a_t \in \mathcal{A}\}$. Like most DRL algorithms, our proposed solution includes training and implementation stages, as shown in Algorithm \ref{my_algo}.

Initially, all hyper-parameters (given in Table \ref{table: all_settings}) are initialized and the network environment is created. The main loop iterates based on the number of experimental steps, \mbox{i.e., the total number of time frames.} In each time frame, the inner loop iterates according to the number of time slots in that frame. Several parameters are set for each time slot, including the Rayleigh distribution parameter ($\lambda$), the number of channel realizations ($\Lambda$), and noise ($\tt NC$). Similarly, random values are generated for the transmit power of all UEs using a uniform distribution, where $P_{\mathtt{UE}_N}$ represents the transmit power of the pUEs, $P_{\tt J}$ is the transmit power of the jammers, and $P_{\tt iUE}$ is the transmit power of the iUE. Next, random values are generated for the channel coefficients, where $h_{\mathtt{UE}_N}$, $h_{\tt J}$ and $h_{\tt iUE}$ are the channel coefficient and path loss exponent for the pUEs, the jammers, and the iUE, respectively.

Next, the algorithm computes the decoding vector $\tt {VN_{UE_{N}}}$ using \eqref{mf_r1} and the SINR for all the UEs in the network using \eqref{mf_n}, where $\mathtt{SINR}_{\tt {UE_N}}^{\tt MF-SIC}$, $\mathtt{SINR}_{\tt {J}}^{\tt MF-SIC}$$\mathtt{SINR}_{\tt {iUE}}^{\tt MF-SIC}$ denote the SINR values of the pUEs, the jammers, and the iUE, respectively. Finally, the transmission rate of all UEs in the network is computed using \eqref{eq: physical_rate_slot}. At this stage, the physical-layer computations at each time slot are complete.

Next, the RL part starts, in which the iUE observes the initial state of the wireless network denoted by $s_0$. In our considered problem, the state comprises the UEs' transmission schedules and the jammers' jamming behavior in the previous time slots. The iUE takes action ($\tt dispatch$ or $\tt hold$) in the current time slot based on the current state of the network in accordance with the $\epsilon$-greedy approach. In the iUE's initial interactions with the network, it prefers exploratory actions more, and thus it chooses actions randomly. However, as time progresses, it prefers to exploit its up-to-date network knowledge. This idea is implemented using exponentially decaying $\epsilon$ with passing time slots. Meanwhile, $C^{[k]}$ broadcasts $ACK^i_{t+1}$, which confirms the channel status and $r_{t+1}$ for the iUE. The iUE stores the current interaction ($s_t, a_t, r_{t+1}, s_{t+1}$) to $\tt EB$. Further in the process, the iUE randomly samples $
\tt {N_{BE}}$ experiences. For each experience, the ResNet computes $Q(s_t, a_t)$ using the prediction network and $Q_{\tt target}(s_{t+1}, a^{'})$ using the target network, where $a^{'}$ is the action that maximizes the Q-value in the next state. Furthermore, the ResNet's parameters are updated using \eqref{eq: update_theta_target} and \eqref{eq: theta_update_final}.
Lastly, when training is complete, i.e., the $\mathcal{L}_{\tt MSE}$ in \eqref{eq:loss_with_e_replay} is negligible, the iUE uses $\pi^*(s)$ to choose optimal transmission actions. 
\subsection{Optimal Policy}
\begin{algorithm}[t]
\caption{A robust DRL-based channel access for varying frame sizes and multi-user settings }\label{my_algo}
 {\bf Input: } $\tt N_{\tt UEs}, N_{\tt J}, \mathcal{A} , S, \epsilon, \gamma, \tau, \tt EB, \tt N_{BE}, \vec{\theta}^{pred}, \vec{\theta}^{target}$\\
{\bf Initialize: } $s_0 \in \mathcal{S}, a_0 \in \mathcal{A}$, set $s_{t+1}= 0, r_{t+1}=0$ \\
\nl $\tt env \leftarrow \tt Network(p, \alpha , s_0, \epsilon, \epsilon_{min}, \epsilon_{decay}, \gamma, \tau, N_{\tt UEs}, N_{\tt J})$ \\
\For {$T: \{T_1, T_2, \cdots, T_f, \cdots, T_F\}$}{
 \For{$t^s_f \in T_f= \{t^1_f,t^2_f,\cdots\, t^S_f\}$} {
  Set $\lambda=1, \tt PLE=1, \Lambda=1$, \\
         $NC \sim \mathcal{U}(2\,\tt dBm,\,5\,\tt dBm)$\\
          $[P_{\tt UE_1}, \cdots, P_{\tt UE_N}, P_{\tt J}, P_{\tt {iUE}}] \sim \mathcal{U}(20\,\tt dBm,\,25\,\tt dBm)$ \\
          Compute $[h_{\tt UE_n}, h_{\tt J_m}, h_{\tt {iUE}}]$ using $\lambda$ \& $\Lambda$ $\forall n\in N,\, \forall m\in M$. \\ 
         
          Calculate $\tt V_{\tt UE_n}$ via \eqref{mf_r1} $\forall n\in N$\\
          Obtain $\mathtt{SINR}_{\tt UE_n}^{\tt MF-SIC}$ via \eqref{mf_n} $\forall n\in N$\\
          Calculate $C_{\tt UE_n}$ via \eqref{eq: physical_rate_slot} $\forall n\in N$\\
          Observe $s_0$ \& update $a_t \leftarrow \epsilon$-greedy([$s_0$]) \\ 
          Get $r_{t+1}, \mathrm{ACK}^{i}_{t+1}, s_{t+1}$ from $\tt env.step$ ([$a_t$])\\
          Deposit $(s_t, a_t, r_{t+1}, s_{t+1})$ to $\tt EB$\\
          \If{$t\%\tau==0$}
                {
                 Update $\vec{\theta}^{\tt target}$ using \eqref{eq: update_theta_target}
                }
          Sample $\tt{E^{N_{BE}}}$ experiences  [$s_t^{\tt N_{BE}}$, $a_t^{\tt N_{BE}}$, $r_{t+1}^{\tt N_{BE}}$, $s_{t+1}^{\tt N_{BE}}]$ \\
         
          \For {each $e \in \tt{E^{N_{BE}}}$ = $(s_t, a_t, r_{t+1}, s_{t+1})$ }
             {
              Compute $Q(s_t, a_t; \theta^{\tt pred})$  \\
              Compute $Q (s_{t+1}, a';\theta^{\tt target})$ \\
              Compute $\mathcal{L}_{\tt MSE_{B}}(\vec{\theta})$ \\ 
              Update $\vec{\theta}^{\tt pred}$ using \eqref{eq: theta_update_final} \\
              Train iUE using $(s_t, Q(s_t, a_t; \vec{\theta}^{\tt pred}), \tt {N_{BE}})$
             }
          Update $s_{t} \leftarrow s_{t+1}$\\
          Optimal $\theta^{\tt pred} \rightarrow \pi^*(s)$
         \\
          Choose $a^*_t \leftarrow \pi^*(s)$}
 } 
\end{algorithm}
\setlength{\aboverulesep}{0pt}
\setlength{\belowrulesep}{0pt}
\begin{table*}[t]
\centering	
 \caption{The settings for network simulation, Neural network's hyper-parameters, and iUE training.}
    \label{table: all_settings}

	\begin{tabular}{@{}llll@{}}
		\toprule
		\multicolumn{2}{l}{\hspace{-3mm} \textbf{Network Simulation Settings}} & \multicolumn{2}{l}{\textbf{Neural network's hyper-parameters}} \\ \midrule
 Number of cells & 7 & Optimizer  & Adam\\
		Number of pUEs  & $10\; \text{to}\;35$ & Loss Type       & Mean squared Error (MSE)   \\
		Number of Jammers  & $2\; \text{to}\;5$ & Batch size       & $32$ \\
		Number of iUE's per cell & 1 & Training epochs       & $70$ \\
		Rayleigh distribution parameter              & 1/$\sqrt{2}$ per dimension                                     & Learning rate      & $0.001$ (default)\\
		Number of channel realization      & $100$  &  Discount factor       & $0.9$ \\
	Transmit power of pUEs [dBm]       & $\text{Uniform}[20,25]$                         & $\epsilon$ in $\epsilon$-greedy     & $[1,0.001]$                          \\
		Transmit power of jammers [dBm]     & $\text{Uniform}[20,30]$                      & $\theta^{\tt target}$ update parameter & $10$ \\
	Noise variance             & 1                  & Experience buffer capacity   & $1000$ \\ 
 
		\cmidrule(l){3-4} 
Frame size	     &       $5\; \text{to}\;30$ slots                  & \multicolumn{2}{l}{\scriptsize \textbf{ResNet Model Settings}}                           \\ \cmidrule(l){3-4} 
Training Episodes	            &   $3000$                          & Input layer        & shape $= (\text{state\_size}= 39)$       \\

		           \cmidrule(r){1-2}
		\multicolumn{2}{l}{\hspace{-3mm} \textbf{OS and other packages}}                           & Dense layer       & units $= 32$, activation $=\tt ReLU$ \\
  \cmidrule(r){1-2}
		 Operating system         &        Ubuntu   $20.04.5$ LTS               & Residual Blocks ($\times 5$ blocks) & units $= 32$, activation $=\tt ReLU$                                                 \\
		           Python           & $3.7.13$                            & Flatten       & $\tt Flatten()$ \\
		            TensorFlow               & $2.9.1$                          & Dense layer & units $= 128$, activation $=\tt ReLU$         \\ 
        
		Keras                        & $2.9.0$                & Dense layer       & units $= 128$, activation $=\tt ReLU$ \\ 
		  GPU     &     GeForce RTX 2080                                  & Output layer & units $=(\mathcal{A} \dot=\{a_t: i = 1, 2\})\times (\tt N_{\tt UEs} + N_{\tt J})$                    \\
    \bottomrule
	\end{tabular}
\end{table*}

An optimal policy $\pi^*(s)$ is a policy that yields the highest expected reward from each state and guides the agent towards achieving its goal. Thus, $\pi^*(s)$ satisfies $Q_{\pi^*}(s_t, a_t) \geq Q_{\pi}(s_t, a_t)$ for all possible $(s_t, a_t) \in \mathcal{S} \times \mathcal{A}$. More specifically, $\pi^*(s)$ can be derived from the optimal action-value function $Q_{\pi^*}(s_t, a_t)$ by choosing the action that maximizes the action-value function for each state-action pair $\pi^*(s_t) ={\argmax_{a_t}}{Q}_{\pi^*}(s_t, a_t)$. Once training is complete, the agent can follow $\pi^*(s)$ to choose optimal actions. 
Fig \ref{fig: JTMAC_optimal_policy} demonstrates how the iUE leverages $\pi^*(s)$ to choose optimal transmission actions $a^*(s)$ in each time slot when the network has 2 pUEs and 1 jammer.
\begin{figure}[!h]
\begin{align*} 
	&T^{(s)}_f \!\!=\!\! 
	\begin{Bmatrix}[l|l] 
s_1 \!\!=\!\! [a^{[k]}_{\tt UE_1}\!\!=\!\!1, a^{[k]}_{\tt UE_2}\!\!=\!\!0, a^{[k]}_{\tt J_1}\!\!=\!\!1]& [\pi^*(s)] \to a^*(s)\!\!=\!\![0] \\
s_2 \!\!=\!\! [a^{[k]}_{\tt UE_1}\!\!=\!\!1, a^{[k]}_{\tt UE_2}\!\!=\!\!1, a^{[k]}_{\tt J_1}\!\!=\!\!0]& [\pi^*(s)] \to a^*(s)\!\!=\!\![0]\\
s_3 \!\!=\!\! [a^{[k]}_{\tt UE_1}\!\!=\!\!1, a^{[k]}_{\tt UE_2}\!\!=\!\!1, a^{[k]}_{\tt J_1}\!\!=\!\!1]& [\pi^*(s)] \to a^*(s)\!\!=\!\![0] \\ 
s_4 \!\!=\!\! [a^{[k]}_{\tt UE_1}\!\!=\!\!0, a^{[k]}_{\tt UE_2}\!\!=\!\!0, a^{[k]}_{\tt J_1}\!\!=\!\!0]& [\pi^*(s)] \to a^*(s)\!\!=\!\![1] \\
s_5 \!\!=\!\! [a^{[k]}_{\tt UE_1}\!\!=\!\!0, a^{[k]}_{\tt UE_2}\!\!=\!\!0, a^{[k]}_{\tt J_1}\!\!=\!\!1]& [\pi^*(s)] \to a^*(s)\!\!=\!\![0] \\
	\end{Bmatrix}
\end{align*}
\caption{Extraction of $a^*$ from [$\pi^*$]}
\label{fig: JTMAC_optimal_policy}
\end{figure}
\subsection{Computational Complexity and Resource Consumption Analysis}
We now present a comprehensive analysis of the computational complexity and resource consumption of the proposed ResNet-based DQN algorithm. This analysis encompasses both the processing-related computational demands and storage requirements associated with the experience replay buffer, offering a comprehensive understanding of the system's efficiency and scalability.

State space $\mathcal{S}$ encompasses the number of pUEs, jammers, and the channel's status, represented as a vector of length $N_{\tt{pUEs}} + N_{\tt{J}} + \mathcal{C}$, where $\mathcal{C}$ denotes the channel's status (i.e., unused, occupied, or being jammed) in a single time slot. Therefore, complexity of managing state space scales linearly with the number of coexisting pUEs and jammers, expressed as $\mathcal{O}(N_{\tt{pUEs}} + N_{\tt{J}} + \mathcal{C})$, thus ensuring that the system remains scalable with an increase of network size.

In conventional fully connected DQN architectures, computational complexity for a forward pass is characterized by $\mathcal{O}(L \cdot H^2)$, where $L$ is the number of layers and $H$ is the number of neurons per layer. This quadratic dependence on $H$ can lead to substantial computational overhead, particularly with an increase of network width.

Our proposed ResNet-based DQN approach mitigates this issue by balancing the depth-versus-width trade-off. Using deeper networks with narrower layers (i.e., fewer neurons per layer), the model retains its ability to learn complex representations while minimizing computational cost. This approach is supported by previous research such as \cite{hoefler2021sparsity} and \cite{zhou2020go}, which demonstrated that deeper networks with narrower layers can achieve superior performance to shallower, wider networks while reducing computational complexity.

The per-layer complexity in ResNet architecture is expressed as $\mathcal{O}(H_{r,m}^{\tt{in}} \cdot H_{r,m}^{\tt{out}})$, where $H_{r,m}^{\tt{in}}$ and $H_{r,m}^{\tt{out}}$ denote the input and output dimensions of the $m$-th layer in the $r$-th residual block, respectively. Since the network is designed with narrower layers (i.e., $H_{r,m}^{\tt{in}}$ and $H_{r,m}^{\tt{out}}$ are relatively small), the per-layer computational cost is substantially reduced as compared to traditional DQNs with wider layers.

The total computational complexity of the ResNet-based architecture, consisting of $R$ residual blocks with $M$ layers each, is computed by summing the complexities across all layers, and can be expressed as
\begin{align} \mathcal{O}\bigg(\!\!(N_{\tt{pUEs}} \!\!+\!\! N_{\tt{J}} \!+\! \mathcal{C}) \!\cdot\! H_0 \!+\! \sum_{r=1}^{R}\! \sum_{m=1}^{M} \!H_{r,m}^{\tt{in}}\! H_{r,m}^{\tt{out}} \!+\! H_f \!\cdot\! |\mathcal{A}| \bigg), \end{align}
where $H_0$ is the number of neurons in the first hidden layer following the input layer, $H_f$ is the number of neurons in the last hidden layer before the output layer, and $|\mathcal{A}|$ is the number of possible actions. Specifically, the term $(N_{\tt{pUEs}} + N_{\tt{J}} + \mathcal{C}) \cdot H_0$ represents complexity of the input layer transformation, as each element of the input vector is connected to $H_0$ neurons in the first hidden layer. Moreover, the double summation $\sum_{r=1}^{R} \sum_{m=1}^{M} H_{r,m}^{\tt{in}} H_{r,m}^{\tt{out}}$ accounts for the complexity of all layers within the residual blocks, where each layer's complexity is determined by the product of its input and output dimensions. Finally, the term $H_f \cdot |\mathcal{A}|$ represents the complexity of the output layer, which maps from the last hidden layer to the Q-values for each action.

During training, updating the Q-values involves processing mini-batches of experience from the replay buffer. The computational complexity for processing a mini-batch of size $N_{\tt{BE}}$ is the product of the complexity per sample and the number of samples in the mini-batch. Therefore, the total computational complexity, which captures the total computational cost for both the forward and backward passes during the training of the network on a mini-batch, can be expressed as
\begin{align}\label{comp_all_actions_updated} \mathcal{O} \! \bigg(\!\!\!N_{\tt{BE}} \!\cdot\!\! \big(\!(N_{\tt{pUEs}} \!\!+\!\! N_{\tt{J}} \!\!+\!\! \mathcal{C})\!\!\cdot\!\! H_0 \!\!+\!\! \sum_{r=1}^{R}\! \sum_{m=1}^{M}\!\! H_{r,m}^{\tt{in}} \!H_{r,m}^{\tt{out}} \!\!+\!\! H_f \!\!\cdot\!\! |\mathcal{A}|\! \big)\!\!\! \bigg). \end{align}

In addition to computational complexity, storage complexity of the experience replay buffer must also be considered. The buffer stores $N_{\tt{BE}}$ experiences, each comprising a state, action, reward, and the subsequent state. Therefore, storage complexity is proportional to the buffer size and the dimensionality of the state-action space, which can be expressed as
\begin{align} \mathcal{O}\left( N_{\tt{BE}} \cdot \left( N_{\tt{pUEs}} + N_{\tt{J}} + \mathcal{C} + |\mathcal{A}| \right) \right). \end{align}

This storage requirement ensures that the replay buffer can efficiently store and retrieve experiences without imposing significant memory overhead, especially since $\mathcal{C}$ is a small constant representing the channel status.

By using narrower layers and an optimal balance between depth and width, our ResNet-based DQN significantly reduces computational complexity as compared to traditional fully connected DQN. The depth-versus-width trade-off allows us to maintain representational capacity of the network while reducing the number of parameters and computational overhead. Narrower layers decrease the per-layer computational cost, leading to faster computations and lower memory requirements.

Furthermore, using hardware accelerators such as GPUs can further reduce computational overhead, particularly during training phases. This architectural design minimizes memory consumption and accelerates computation during both training and inference, thereby making the proposed approach a practical and scalable solution for real-world network environments.
\section{Simulation Setup and Performance Evaluation}\label{section: simulation_results}
\subsection{Simulation Setup}
We used Python for all our experimental simulations. The RL framework was simulated using Keras and TensorFlow. Other important information regarding the simulation settings, neural network hyper-parameters, platforms installed, and packages used is provided in Table~\ref{table: all_settings}.
\subsection{Simulation Environment}
\begin{figure}[!t]
\centering
\includegraphics[width=\linewidth]{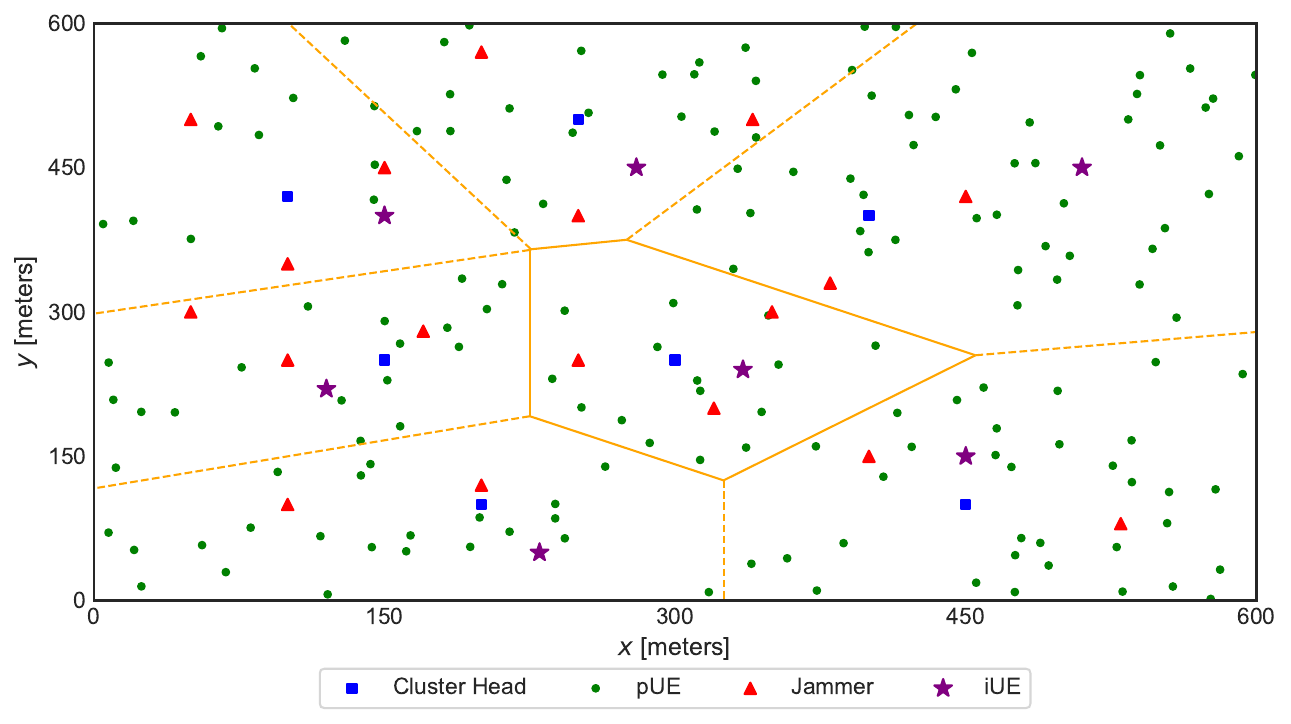}
\caption{Multi-cell smart grid network deployment, with pUEs, jammers, and iUEs randomly distributed across cells, simulating realistic interference and diverse access protocols.}
\label{fig: deploy_scene}
\end{figure}

To evaluate performance of our proposed algorithm in a realistic setting, we conducted extensive simulations based on an smart grid infrastructure network application, as shown in Fig.~\ref{fig: deploy_scene}. This environment consisted of a multi-cell network with multiple pUEs (smart meters) randomly deployed across cells and an iUE deployed in each cell. In addition, several random jammers were introduced into the network to simulate malicious jamming.
The scenario modeled a multi-cell smart grid network where pUEs, with diverse transmission schedules, periodically transmit data to their respective $\tt CH$s for monitoring and billing purposes. Likewise, jammers with different jamming patterns aimed to disrupt this legitimate communication. To simulate a more realistic network environment, we introduced heterogeneity by modeling the use of different access protocols for pUEs. In this context, the iUE in each cell operated as a fault detector, opportunistically adapting its transmissions (e.g., reporting equipment failures, network connectivity issues, power outages) to the $\tt CH$, based on the learned transmission patterns of pUEs, while navigating jamming, intra-cell and inter-cell interference, and varying channel conditions. Accordingly, this setup created a highly challenging environment, allowing us to evaluate the effectiveness of our proposed algorithm in handling the coexistence of an iUE with legitimate pUEs and jammers in a particular cell under dynamic channel conditions and both inter-cell and intra-cell interference.

The simulation environment consisted of 7 cells, with each cell containing 2 to 5 jammers, 10 to 35 pUEs, 1 $\tt CH$, and 1 iUE. All devices communicated with their respective $\tt CH$s through a time-slotted uplink channel. We used a Rayleigh fading channel model to capture multi-path propagation and signal attenuation. In line with our system model, we ignored large-scale fading and focused solely on small-scale fading effects. Simulation parameters used in our experiments are summarized in Table~\ref{table: all_settings}.
\begin{figure*}[!t]
\centering
\includegraphics[width=.9\linewidth]{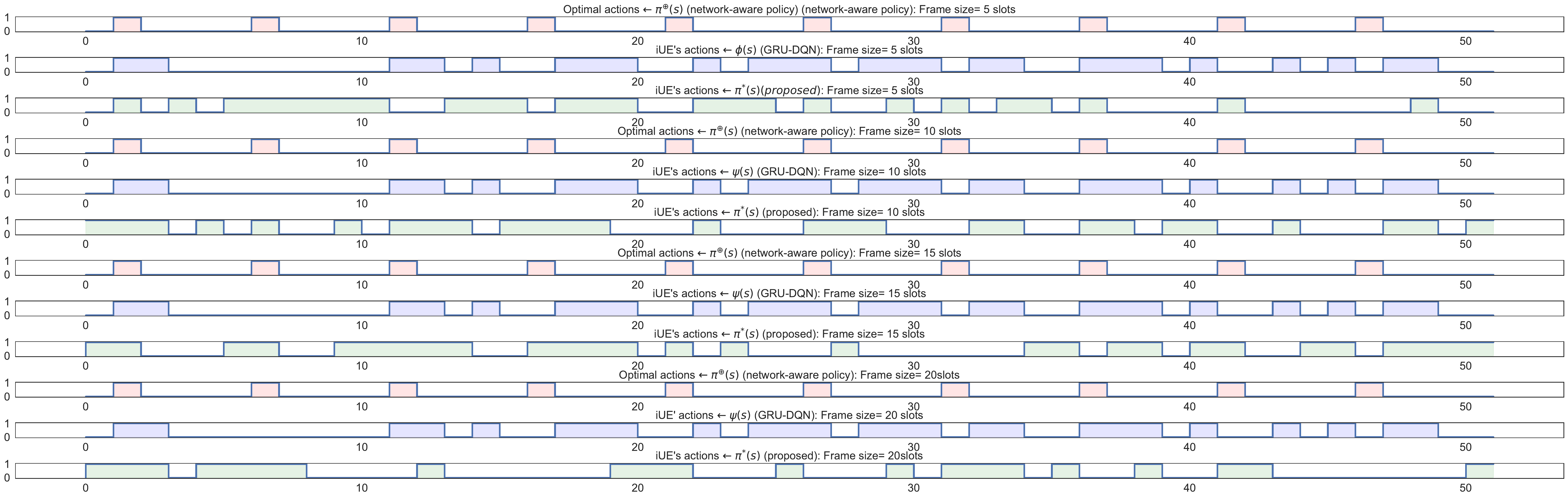}
\caption{iUE's transmission actions (before training).}
\label{fig:agent_action_unlearned}
\includegraphics[width=.9\linewidth]{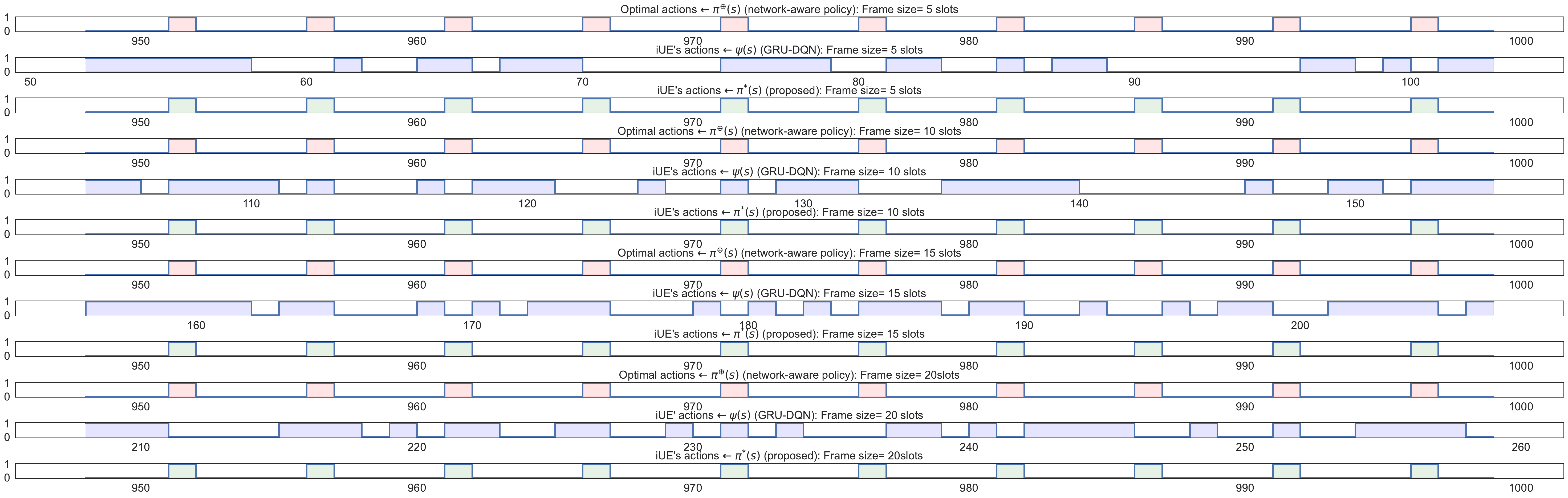}
\caption{iUE's transmission actions (after training).}
\label{fig:agent_action_learned}
\end{figure*}

\subsection{Benchmarks}
The proposed model-free learning framework, denoted by $\pi^*(s)$, enables the iUE to learn an optimal transmission policy without explicit knowledge of behaviors of coexisting pUEs and jammers. We evaluate our approach by comparing it against the following three benchmarks.
\begin{itemize}
    \item \textbf{FC-DQN}: This benchmark is a policy denoted as $\psi(s)$, learned by the iUE using a fully connected deep Q-network (FC-DQN) as proposed in \cite{ali2023defeating}. The FC-DQN approximates the Q-function using a feed forward neural network architecture without recurrent connections, processing each state individually.
    \item \textbf{GRU-DQN}: This benchmark is a policy denoted as $\phi(s)$, which uses a gated recurrent unit (GRU)-based deep Q-network as proposed in \cite{ye2021multi}. The GRU-DQN incorporates temporal dependencies through its recurrent structure, allowing it to model sequential patterns in the state transitions.
    \item \textbf{Network-aware policy}: An optimal reference is the policy of a network-aware UE (nAUE) denoted by $\pi^{\oplus}(s)$. The nAUE possesses accurate knowledge of the pUEs' schedules and jammers' behavior before the start of the simulation. Accordingly, using $\pi^{\oplus}(s)$, the nAUE always transmits in the unused time slots, starting from state $s_0$. Consequently, it receives the highest reward at each time slot for avoiding collisions with pUEs' packets and jammers' jamming, irrespective of frame size and combination of coexisting pUEs and jammers. Accordingly, $\pi^{\oplus}(s)$ serves as an optimal reference point, representing the theoretical upper limit of performance that any UE can achieve in this environment.
\end{itemize}

\subsection{Performance Evaluation}\label{sec:Performance_evaluation}
In this section, we evaluate performance of our proposed solution. We start by illustrating the optimal behavior of the iUE based on its learned policy and highlight the actions it selected. We then examine the iUE’s learning curve under various learning frameworks and explore the impact of frame size variations on convergence. This is followed by the analysis of how changing the number of pUEs and jammers affects the iUE’s learning and of the iUE’s losses during training. Finally, for a comprehensive analysis of the iUE’s learning performance, we present the SCLAR plot.
\subsubsection{Optimal iUE Actions}
Fig.~\ref{fig:agent_action_unlearned} shows the results of our analysis of the iUE's actions in its initial interactions with the network for different frame sizes when following policies $\pi^{*}(s)$ and $\phi(s)$. For the sake of comparison, we present the actions of an iUE following $\phi(s)$ and optimal actions (derived from $\pi^{\oplus}(s)$). As can be seen in Fig.~\ref{fig:agent_action_unlearned}, irrespective of the frame size and policy, the iUE initially makes random transmission actions. This is so because the iUE has not yet learned the network dynamics, which results in packet loss, low channel utilization, and adversely affects the network SCLAR. 

However, as shown in Fig.~\ref{fig:agent_action_learned}, in its final interactions with the network, the iUE following $\pi^{*}(s)$ consistently outperforms an iUE that uses $\phi(s)$ across all frame sizes. More specifically, by following $\pi^{*}(s)$ derived from our proposed solution, the iUE judiciously chooses optimal transmission actions.
\subsubsection{Different Learning Methods and Frame Sizes}

\begin{figure}[!t]
\centering
\includegraphics[width=\linewidth]{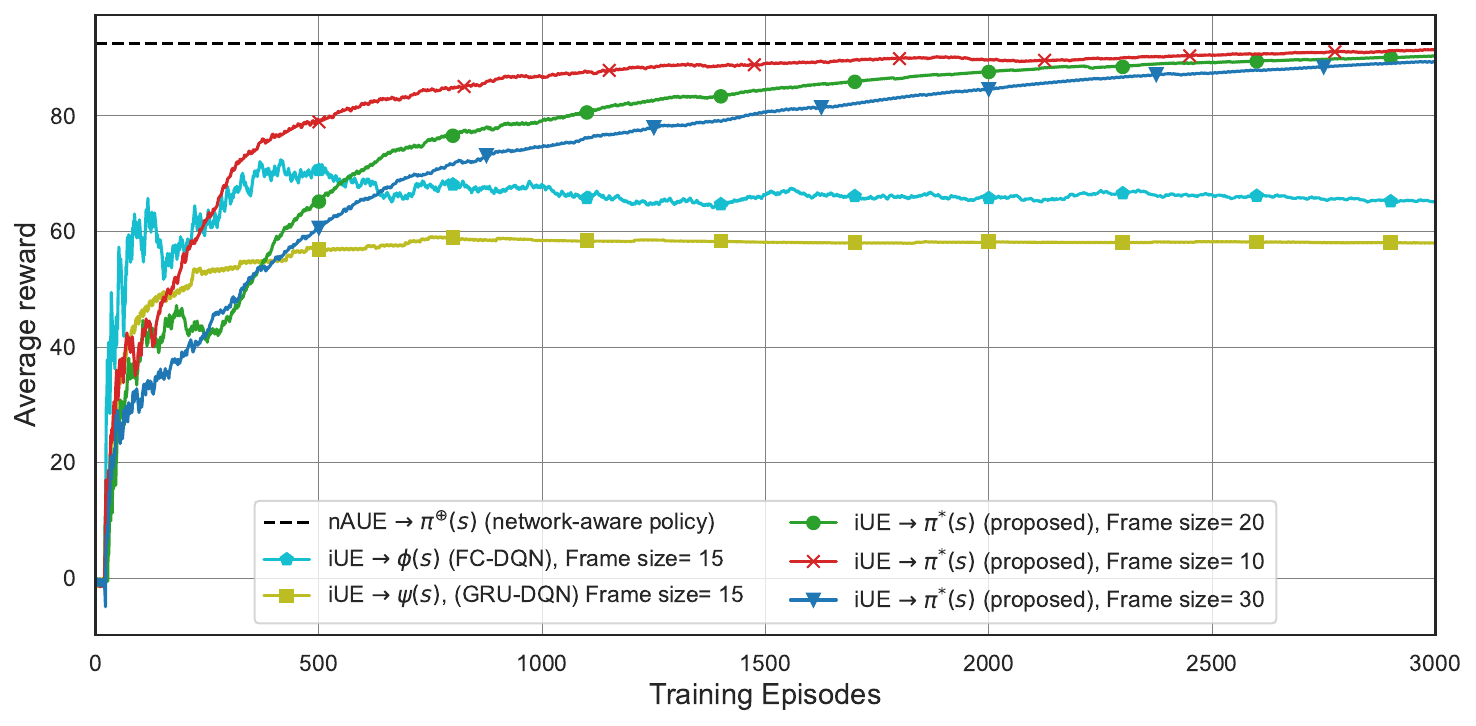}
\caption{Agent's learning curve with different learning methods and frame sizes.}
\label{fig:learning_curve}
\end{figure}

Fig.~\ref{fig:learning_curve} illustrates the iUE's learning curves generated by using various learning methods in a network configuration consisting of 7 cells, each with 20 pUEs and 3 jammers. Apart from dynamicity in the channel, this configuration introduces additional complexities, such as inter-cell and intra-cell interference. These learning curves represent the iUE's average cumulative rewards across training episodes.

The dashed line in Fig.~\ref{fig:learning_curve} shows the optimal average reward of nAUE using the network-aware policy $\pi^{\oplus}(s)$. Fig.~\ref{fig:learning_curve} clearly indicates that the iUE's learning curve does not converge to optimal rewards when using $\phi(s)$. This outcome is due to the GRU's limitations in handling temporal dependencies in complex network dynamics, which results in sub-optimal action selection and lower convergence values.

Conversely, with $\psi(s)$, the learning curve converges to a sub-optimal average reward value. While FC-DQN can capture schedules of coexisting pUEs and jammers, it experiences difficulties in comprehending complexities arising from dynamic channel conditions and interference. Therefore, when using $\psi(s)$, the iUE gets trapped in a local maximum, which results in transmission actions based on flawed knowledge. 

However, when employing $\pi^*(s)$ derived from our proposed framework, the iUE consistently converges to near-optimal reward values across different frame sizes, which demonstrates its long-term stability and effective adaptation to complex network dynamics. This finding clearly shows superior feature extraction capabilities and learning efficiency of the proposed method, enabling the iUE to effectively adapt to the complex dynamics of the network environment.
 \subsubsection{Varying Number of pUEs and Jammers}
\begin{figure}[!t]
\centering
\includegraphics[width=0.95\linewidth]{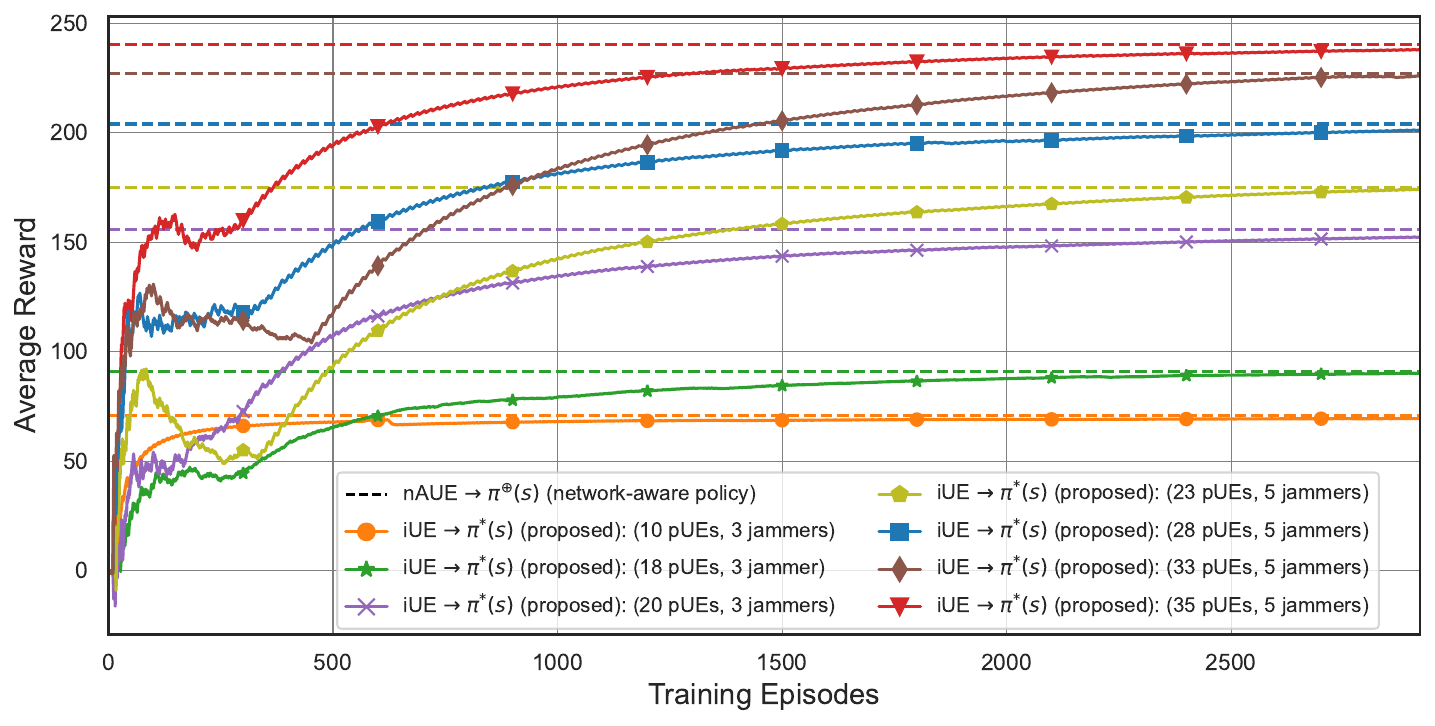}
\caption{Agent's learning curve with varying pUEs and jammers.}
\label{fig: varying_users}
\end{figure}

 Fig. \ref{fig: varying_users} plots the iUE's learning curve when different combinations of pUEs and jammers coexist in the network, with the frame size of 30. The dashed lines shows the optimal reference when iUE follows $\pi^{\oplus}(s)$ and coexists with the respective number of pUEs and jammers. As can be seen in Fig.~\ref{fig: varying_users}, with 10 pUEs and 3 jammers, the iUE can quickly learn the network dynamics and converge to the near-optimal reward. However, with an increase in the number of pUEs and jammers, convergence time increases; yet, the iUE ultimately converges to the near-optimal reward. Moreover, since the reward function in \eqref{eq: reward_function} includes the sum of the utility values of all coexisting pUEs, with an increase in the number of pUEs, the iUE's average reward also increases. These attributes demonstrate the self-adaptive capability of the proposed framework, thus enabling the iUE to manage an increasing number of coexisting devices and effectively adapt to network dynamics.
\subsubsection{Training Losses}
\begin{figure}[!t]
\centering
\includegraphics[width=0.98\linewidth]{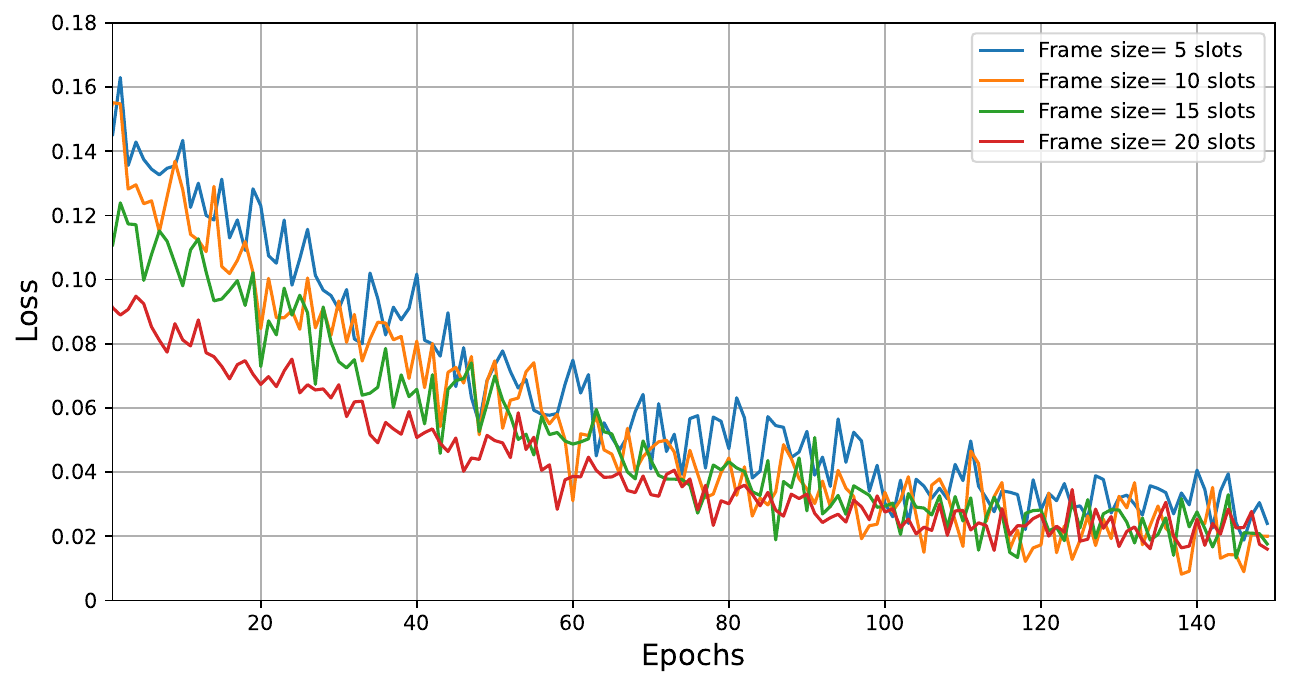}
\caption{Epoch Loss in final episode when iUE uses proposed $\pi^{*}(s)$.}
\label{fig:epoch_loss}
\end{figure}

Fig.~\ref{fig:epoch_loss} plots epoch loss against the number of training epochs under different frame sizes. As can be seen in Fig.~\ref{fig:epoch_loss}, the ResNet-based iUE successfully reduces the amount of loss in each successive epoch for all considered frame sizes. The loss reduction observed suggests that the ResNet improves its prediction ability with every passing epoch regardless of frame size, thereby improving performance for all training episodes.
\begin{figure}[!t]
\centering
\includegraphics[width=\linewidth]{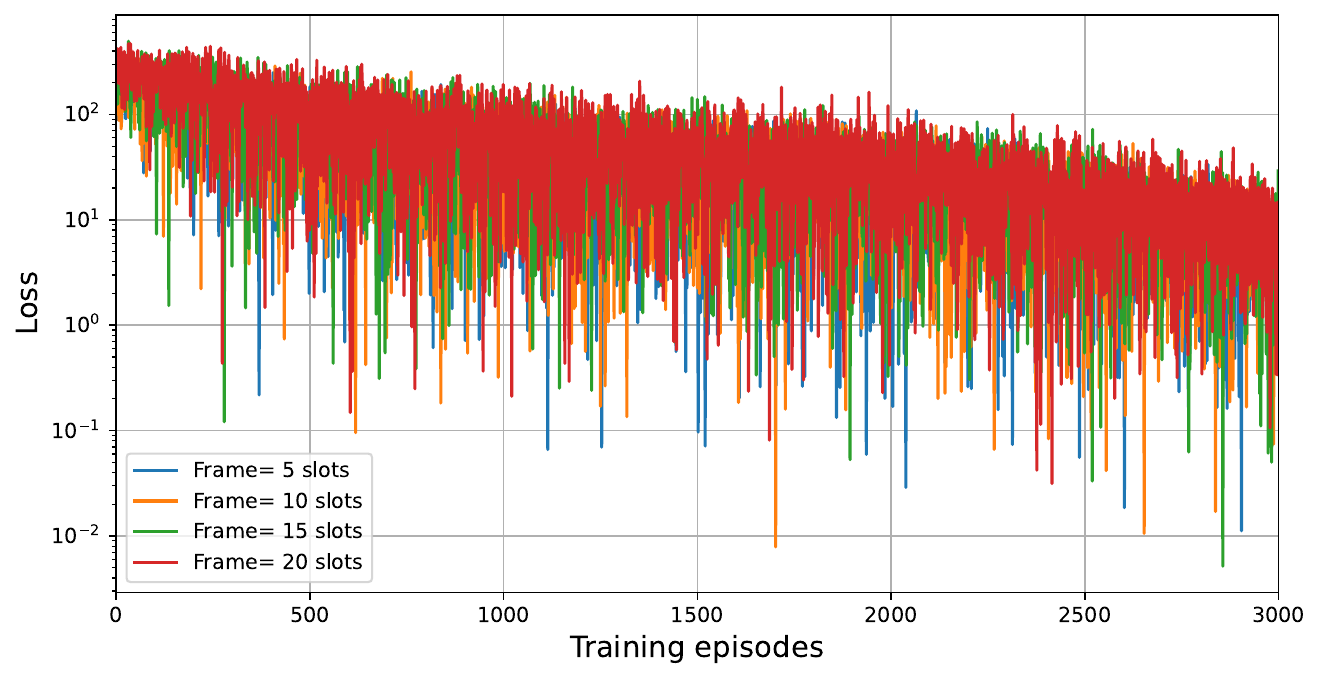}
\caption{Training losses when iUE uses proposed $\pi^{*}(s)$.}
\label{fig: Loss_last_value_all_episodes}
\end{figure}

Fig.~\ref{fig: Loss_last_value_all_episodes} shows the iUE's training loss over the training episodes. As can be seen in the figure, the iUE successfully reduces training loss, which results in an accurate prediction of $Q(s_t, a_t; \vec{\theta}^{\tt pred})$ regardless of frame size. The observed reduction in training loss highlights the iUE's capability for ongoing learning and adaptation, which reinforces its effectiveness in handling diverse network scenarios and maintaining high performance throughout different training phases.
\begin{figure}[t]
\centering
\includegraphics[width=0.97\linewidth]{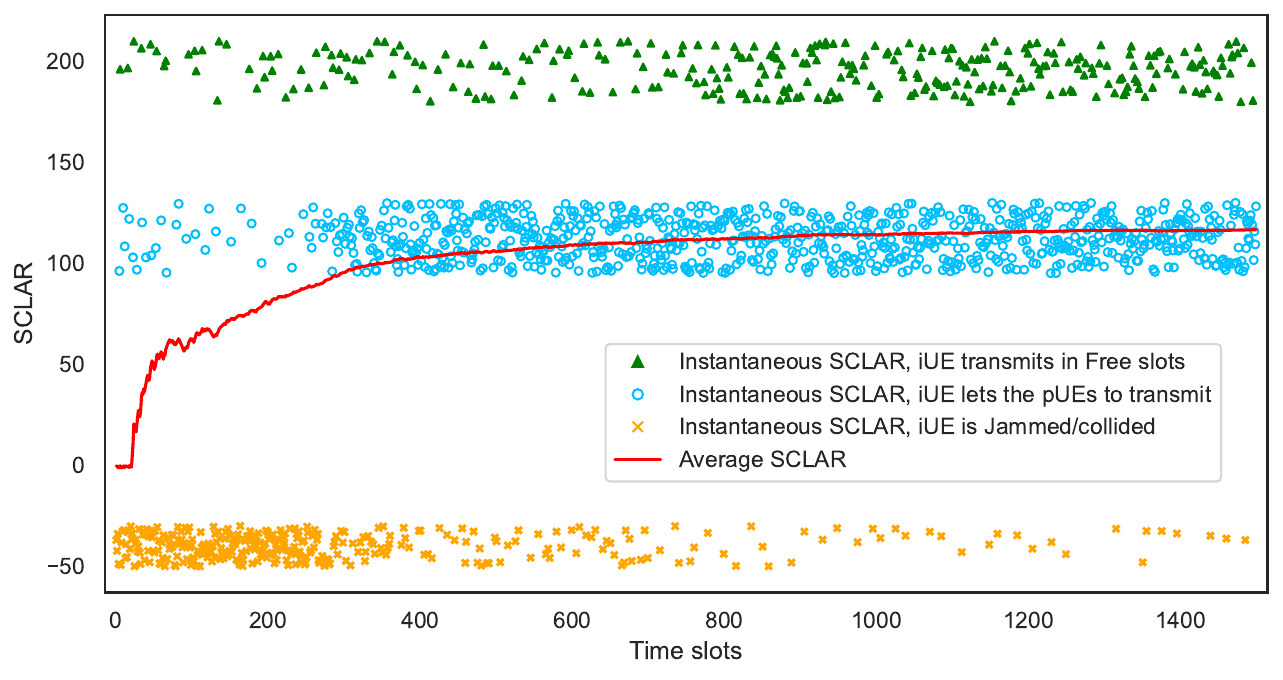}
\caption{SCLAR of the considered network when iUE uses proposed $\pi^{*}(s)$.}
\label{fig:SCLAR}
\end{figure}
\subsubsection{Network's SCLAR}
In Fig.~\ref{fig:SCLAR}, the network's performance is evaluated by examining its instantaneous and average SCLAR across time slots. Fig.~\ref{fig:SCLAR} shows the average SCLAR with a red line, when the frame size is set to 20, and the network has 20 pUEs and 3 jammers. The colored dots represent the instantaneous SCLAR at each time slot. 

Initially, the iUE takes exploratory actions, which results in colliding with coexisting pUEs, wasting unused time slots, and being jammed by the jammers, all of which result in a lower SCLAR. However, as training progresses, the iUE gains more knowledge by learning the pUEs' transmission schedule and jammers' jamming patterns. As a result of such learning, the iUE chooses only those transmission actions that lead to a higher SCLAR in each time slot. Said differently, the iUE transmits packets only in unused time slots and refrains from transmitting when coexisting pUEs are transmitting or when the jammers are active. As demonstrated by the results, the iUE can enhance the network's SCLAR by learning its dynamics after only a few training episodes.

\section{Conclusion}\label{section: conclusions}
In this paper, we presented a DRL-based iUE in a quasi-static wireless network that learns a robust transmission strategy to access a time-slotted uplink channel and maximize the network's SCLAR in the presence of several pUEs and malicious jammers under realistic uplink channel dynamics. The proposed approach was demonstrated to have self-adaptive capabilities, optimizing channel access by continuously learning from its environment to reduce collisions with coexisting pUEs and evade jammers. The simulation results revealed that the iUE can adapt to changing channel conditions and effectively coexist with other pUEs, even in the presence of dynamic interference and jamming attacks. Moreover, the iUE was found to exhibit long-term stability in maintaining high network performance by intelligently selecting transmission slots, ensuring consistent improvements in SCLAR over time. Once trained, the iUE continuously adapted to evolving network conditions, enabling sustained performance optimization in dynamic environments. In future research, it would be meaningful to explore the application of our proposed approach to highly dynamic and autonomous wireless networks, addressing challenges such as rapid topological changes and scalable learning mechanisms so as to enhance adaptability and performance in large-scale environments.

  \balance
\bibliographystyle{IEEEtran}
\bibliography{references}

\end{document}